\documentclass[10pt,journal,cspaper,compsoc]{IEEEtran}
\usepackage{cite}

\ifCLASSINFOpdf
  \usepackage[pdftex]{graphicx}
\else
  \usepackage[dvips]{graphicx}
\fi

\usepackage{amsmath}

\usepackage{algorithmic}

\usepackage{array}

  \usepackage[caption=false,font=normalsize,labelfont=sf,textfont=sf]{subfig}

\usepackage{balance}

\begin{document}

\def\x{{\mathbf x}}
\def\L{{\cal L}}
\def\eg{\textit{e.g.}}
\def\ie{\textit{i.e.}}
\def\Eg{\textit{E.g.}}
\def\etal{\textit{et al.}}
\def\etc{\textit{etc}}
%
\title{Recognition of Advertisement Emotions with Application to Computational Advertising}
%
%
%

\author{Abhinav Shukla, Shruti Shriya Gullapuram, Harish Katti,\\ Mohan Kankanhalli,~\IEEEmembership{Fellow,~IEEE}, Stefan Winkler,~\IEEEmembership{Fellow,~IEEE}, and Ramanathan Subramanian,~\IEEEmembership{Senior Member,~IEEE}
\vspace{-.4in}

\thanks{Abhinav Shukla is with the Imperial College, London.}
\thanks{Shruti Shriya Gullapuram is with the University of Massachusetts, Amherst.}
\thanks{Harish Katti is with the Indian Institute of Science, Bangalore.}
\thanks{Mohan Kankanhalli is with the National University of Singapore.}
\thanks{Stefan Winkler is with the Advanced Digital Sciences Center, Singapore.}
\thanks{Ramanathan Subramanian is with A*STAR Singapore.}
\thanks{Manuscript in submission.}}

\markboth{IEEE Transactions on Affective Computing, Vol.~??, No.~??, January~1970}{Shukla \MakeLowercase{\textit{et al.}}: Emotion Recognition in Advertisements with Application to Computational Advertising}


\maketitle

\begin{abstract}
Advertisements (ads) often contain	strong affective content to capture viewer attention and convey an effective message to the audience. However, most computational affect recognition (AR) approaches examine ads via the text modality, and only limited work has been devoted to decoding ad emotions from audiovisual or user cues. This work (1) compiles an affective ad dataset capable of evoking coherent emotions across users; (2) explores the efficacy of content-centric convolutional neural network (CNN) features for AR vis-\~a-vis handcrafted audio-visual descriptors; (3) examines user-centric ad AR from Electroencephalogram (EEG) responses acquired during ad-viewing, and (4) demonstrates how better affect predictions facilitate effective computational advertising as determined by a study involving 18 users. Experiments reveal that (a) CNN features outperform audiovisual descriptors for content-centric AR; (b) EEG features are able to encode ad-induced emotions better than content-based features; (c) Multi-task learning performs best among a slew of classification algorithms to achieve optimal AR, and (d) Pursuant to (b), EEG features also enable optimized ad insertion onto streamed video, as compared to content-based or manual insertion techniques in terms of ad memorability and overall user experience.
\end{abstract}

\begin{IEEEkeywords}
Affect Recognition; Advertisements; Content-centric vs. User-centric; Convolutional Neural Networks (CNNs); EEG; Multimodal; Multi-task Learning; Computational Advertising; Human vs. Computational Perception; 
\end{IEEEkeywords}

%
\IEEEpeerreviewmaketitle

\section{Introduction}~\label{sec:intro}
\IEEEPARstart{A}{dvertising} is a pivotal industry in today's digital world, and advertisers showcase their products and services as not only useful, but also highly worthy and rewarding. Emotions play a crucial role in conveying an effective message to viewers, and are known to mediate consumer attitudes towards brands~\cite{Holbrook1984,Holbrook1987,Pham2013}. Emotions are also critical for spreading public health and safety awareness, where certain personal choices are portrayed as beneficial to improving one's quality of life, while others are portrayed as deleterious and possibly fatal. Therefore, the ability to objectively characterize advertisements (ads) in terms of their emotional content has multiple applications-- \eg, inserting appropriate ads at optimal temporal points within a video stream can benefit both advertisers and consumers of video streaming websites such as YouTube~\cite{cavva,Karthik2013}. Subjective experience of pleasantness (\textit{\textbf{valence}}) and emotional intensity (\textit{\textbf{arousal}}) are important affective dimensions~\cite{Russell1980}, and both influence responses to ads in distinct ways \cite{Broach1995}. Specifically, stimulus valence and arousal are known to influence recall of images~\cite{Khosla2013}, movie scenes~\cite{Subramanian2014} and videos~\cite{cavva}. 

While mining of ad emotions is therefore beneficial, only a few works have attempted the same. This is despite the popularity of {{affective computing}} recently, and a multitude of works predicting emotions elicited by image~\cite{katti2010making,maneesh2017acii}, speech~\cite{lee2005toward}, audio~\cite{AAAI17}, music~\cite{Koelstra} and movie~\cite{decaf,subramanian2016ascertain} content. Ad affect characterization is non-trivial as with stimuli such as music and movie clips~\cite{Hanjalic2005,wang2006affective,Koelstra,decaf} as human emotional perception is subjective.  In lieu of detecting of discrete emotion categories such as \textit{joy},\textit{ sorrow} and \textit{disgust}, many affect recognition (AR) works model emotions along the valence (val) and arousal (asl) dimensions~\cite{Russell1980,greenwald1989}. Overall, AR methods are broadly classified as \textit{content-centric} or \textit{user-centric}. \textit{Content-centric} AR characterizes emotions by examining textual, audio and visual cues~\cite{Hanjalic2005,wang2006affective}. In contrast, \textit{user-centric} AR identifies elicited emotions from facial~\cite{joho2011looking} or physiological~\cite{Koelstra,decaf,subramanian2016ascertain,Subramanian2014} measurements acquired from the {user} or multimedia consumer. While enabling a fine-grained examination of transient emotions, user-centered methods may nevertheless suffer from individual subjectivity.

This work expressly studies emotions conveyed by ads, and employs (i) explicit human opinions and (ii) associated content and user-centric measurements (or descriptors) which influence these opinions. Firstly, we examined the efficacy of 100 diverse, carefully curated ads to coherently evoke emotions across viewers. To this end, we examined the affective first impressions of five experts and 23 novice annotators and found that the two groups agreed considerably on the asl and val ratings. Secondly, we explored the utility of Convolutional Neural Networks (CNNs) and domain adaptation for encoding emotional audiovisual (\ie, content-based) features. As the compiled ad dataset is relatively small and insufficient for CNN training, we employed {domain adaptation} to transfer knowledge gained from the large-scale and annotated LIRIS-ACCEDE movie dataset~\cite{baveye2015liris} for decoding ad emotions. Extensive experimentation confirms that CNN descriptors outperform handcrafted audio-visual descriptors proposed in~\cite{Hanjalic2005}, with a substantial improvement observed for val recognition. 

Thirdly, we performed user-centric ad AR from EEG responses compiled from annotators, and found that a three-layer CNN trained on EEG features produced state-of-the-art performance for both asl and val recognition. To our knowledge, this is the first work to perform an explicit comparison of content and user-centric methods for ad AR. In addition, we explored the utility of multi-task learning and feature/decision fusion techniques for asl and val classification. Lastly, we examined if accurate encoding of ad emotions facilitated optimized insertion of ads onto a video stream, as ads contribute to revenue generation of video hosting websites such as \textit{YouTube}. A study with 18 viewers confirmed that insertion of ads identified via EEG-based   emotional relevance maximized ad memorability and viewing  experience while watching the ad-embedded video stream. In summary, we make the following contributions:

\begin{itemize}
\item[1.] This is one of the few works to examine AR in ads, extending findings reported in~\cite{Shukla2017icmi,Shukla2017acm}. It is also the only work to characterize ad emotions in terms of explicit human opinions, and underlying (content-centric) audiovisual plus (user-centric) EEG features. 
\item[2.] We present a carefully curated affective dataset of 100 ads and associated affective ratings. Based on statistical analyses, we note that the ad dataset is capable of evoking coherent emotions across the \textit{expert} and \textit{novice annotator} groups.
\item[3.] We examine the utility of CNN-based transfer learning for AR. We show that CNN features, synthesized by fine-tuning \textit{Places205} Alexnet~\cite{places14} effectively captures emotional audiovisual features. Experiments show that CNN features outperform handcrafted audio-visual descriptors proposed in~\cite{Hanjalic2005}.
\item[4.] We compare and contrast AR achieved with audiovisual and EEG-based CNN features. The EEG-based CNN model best encodes the asl and val attributes. Also, multi-task learning to exploit feature similarities among emotionally similar ads considerably benefits ad AR. Finally, probabilistically fusing estimates of multiple classifiers achieves superior AR than unimodal classifiers.  
\item[5.] We demonstrate how improved AR positively impacts ad memorability and user experience while watching an ad-embedded video stream. To our knowledge, this is one of the few works to demonstrate how improved estimation of ad asl and val scores can positively impact a computational advertising application.
\end{itemize} 

The paper is organized as follows. Section~\ref{RW} reviews related literature, while Section~\ref{ad_set} overviews the compiled ad dataset and the EEG acquisition protocol. Section~\ref{Data_anal} presents the techniques adopted for content and user-centered ad AR, while Section~\ref{ER} discusses AR results. Section~\ref{user_study} describes the user study to establish how improved emotion estimation facilitates computational advertising. Section~\ref{CFW} summarizes the main findings and concludes the paper.

\section{Related Work}~\label{RW}
To position our work with respect to the literature and highlight its novelty, we review the related work examining (a) affect recognition (b) the impact of affective ads on consumer behavior (c) computational advertising.

\subsection{Affect Recognition}
Both \textit{content-centric} and \textit{user-centric} approaches have been proposed to infer emotions evoked by multimedia stimuli. Content-centric approaches~\cite{Hanjalic2005,wang2006affective} predict the likely elicited emotions by examining image, audio and video-based emotion correlates~\cite{Hanjalic2005,vonikakis2017probabilistic,Shukla2017acm}. In contrast, user-centric AR methods~\cite{Koelstra,decaf,subramanian2016ascertain} estimate the stimulus-evoked emotion based on physiological changes observed in viewers (content consumers). Physiological signals indicative of emotions include pupillary dilation~\cite{YadatiMMM2013}, eye-gaze patterns~\cite{Subramanian2014,Tavakoli15} and neural activity~\cite{Koelstra,decaf,Zheng2014}. Both content and user centric methods require labels denoting stimulus emotion, and such labels are compiled from annotators whose affective opinions are deemed \textit{acceptable}~\cite{Raykar2010,YeLNAW17}, given that emotion perception is highly subjective. In this work, we show that a carefully curated set of 100 ads are assigned very similar emotional labels by two independent groups comprising \textit{experts} and \textit{novice annotators}. Emotional attribute (\ie, asl and val) labels for these ads are then predicted via content and user-based methods. User-centered AR is achieved via EEG signals acquired with a wireless and wearable \textit{Emotiv} headset, which is minimally intrusive and facilitates naturalistic user behavior.

\subsection{Emotional impact of ads}
Ad-induced emotions influence consumer behavior significantly~\cite{Holbrook1984,Holbrook1987}. Work described in~\cite{Pham2013} concludes that ad-evoked feelings impact viewers explicitly as well as implicitly, and influence change in user attitudes towards (especially hedonistic) products. While many works have examined the correlation between ad emotions and user behavior, very few works have exploited these findings for developing targeted advertising mechanisms. The only work that incorporates emotional information for modeling context in advertising is CAVVA~\cite{cavva}, where ad-in-video insertion is modeled as a discrete optimization problem based on emotion relevance between video scenes and an inventory of ads. Based on consumer psychology rules, video scenes are matched with ads with respect to asl and val scores to determine (a) the suitable ads for presentation and (b) optimal ad insertion points that would maximize user engagement. 

Two recent and closely related works~\cite{Shukla2017acm, Shukla2017icmi} discuss how efficient affect recognition from ads via deep learning and multi-task learning lead to improved online viewing experience. This work builds on~\cite{Shukla2017acm, Shukla2017icmi} to show via extensive experiments that CNNs best encode emotions from both content and user-centered cues. Also, learning feature similarities among \textit{related} stimuli (\eg, high asl, high val and high asl, low val clips) via multi-task learning (MTL) can benefit AR in a data impoverished setting involving only 100 labeled ad exemplars. MTL achieves excellent AR performance with both audiovisual and EEG descriptors. Finally, probabilistically fusing the estimates of multiple classifiers (decision fusion) is found to improve AR performance over unimodal classification. As a demonstration of how better affect estimation impacts a real-life application, we show how audiovisual and EEG-based CNN models enable optimized insertion of ads onto a video sequence with respect to manual performance via the CAVVA framework~\cite{cavva}. The user study confirms that the EEG-based CNN model, which achieves the best AR performance, also results in maximum ad memorability and the best experience for viewers watching an ad-embedded video stream.

\subsection{Computational advertising}
Exploiting affect recognition models for commercial applications has been a growing trend in recent years. The field of \textit{\textbf{computational advertising}} focuses on presenting contextually relevant ads to multimedia users for commercial benefits, social good or to induce behavioral change. Despite the fact that ads are emotional, computational advertising methods have essentially matched low-level visual and semantic properties between video segments and candidate ads~\cite{videosense} for ad display, ignoring emotional relevance. A paradigm shift in this regard was introduced by the CAVVA framework~\cite{cavva}, which proposed an optimization-based approach to insert ads onto video based on the emotional relevance between the video scenes and candidate ads. CAVVA employed a \textit{content-centric} approach to match video scenes and ads in terms of emotional valence and arousal. However, this could be replaced by an interactive and \textit{user-centric} framework as described in~ \cite{YadatiMMM2013}. We explore the use of both \textit{content-centric} (via audiovisual CNN features) and \textit{user-centric} (via EEG features) methods for formulating an ad-insertion strategy. A user study shows an EEG-based strategy achieves optimal user experience and also performs best with respect to ad memorability. The following section positions our work with respect to the literature.

\subsection{Analysis of related work}
Examination of the literature reveals that (1) AR studies are typically hampered by subjectivity in emotion perception, and a control dataset that can coherently evoke emotions in users is essential for effectively learning content or user-based emotion predictors; (2) Despite the fact that ads are emotional, and that ad emotions significantly impact user behavior, very little effort has been devoted towards incorporating emotional video-ad relevance in a computational advertising framework.  

In this regard, we present the first work to compile a control set of affective ads which elicit concordant opinions from both \textit{experts} and \textit{naive users}. Also, we leverage CNNs for learning both audiovisual and EEG-based emotion predictors. Optimal AR is achieved with a CNN classifier employing EEG features, while CNN-based audiovisual descriptors outperform handcrafted counterparts proposed in~\cite{Hanjalic2005}. Finally, we also demonstrate via a user study how better affect encoding facilitates ad-to-video insertion via the CAVVA mechanism~\cite{cavva} to enhance user viewing experience as well as ad memorability. Details pertaining to our ad dataset are presented below.

\section{Advertisement Dataset}~\label{ad_set}
This section presents details regarding our ad dataset along with the protocol employed for collecting user ratings and EEG responses for user-centric AR.

\subsection{Dataset Description}~\label{DD}
\begin{figure}[t]
\includegraphics[width=0.32\linewidth]{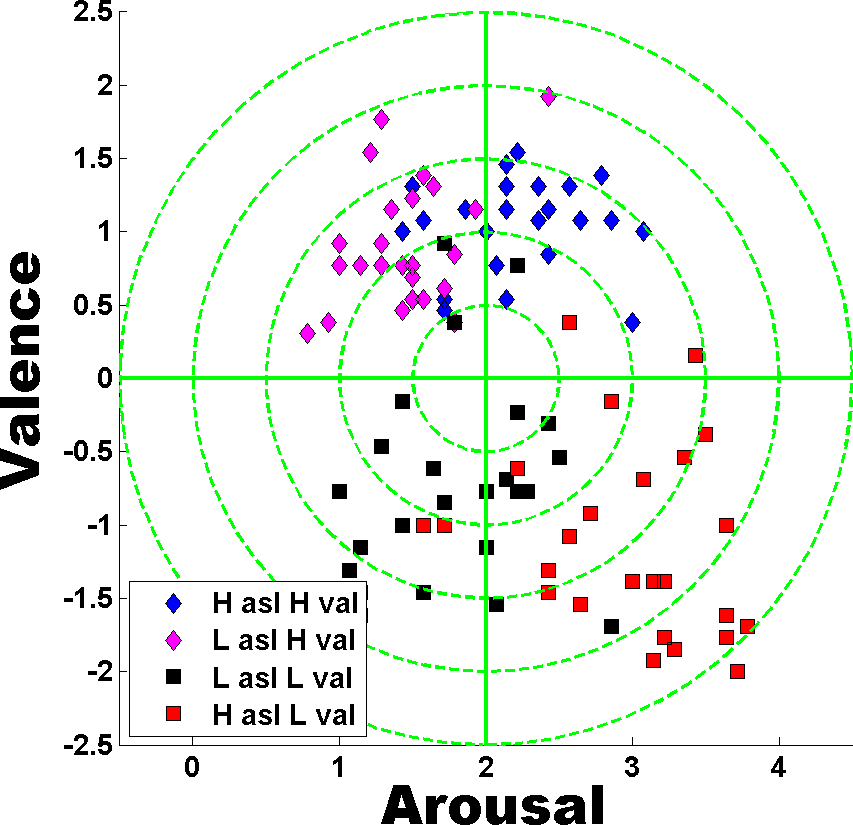}\hfill
\includegraphics[width=0.32\linewidth]{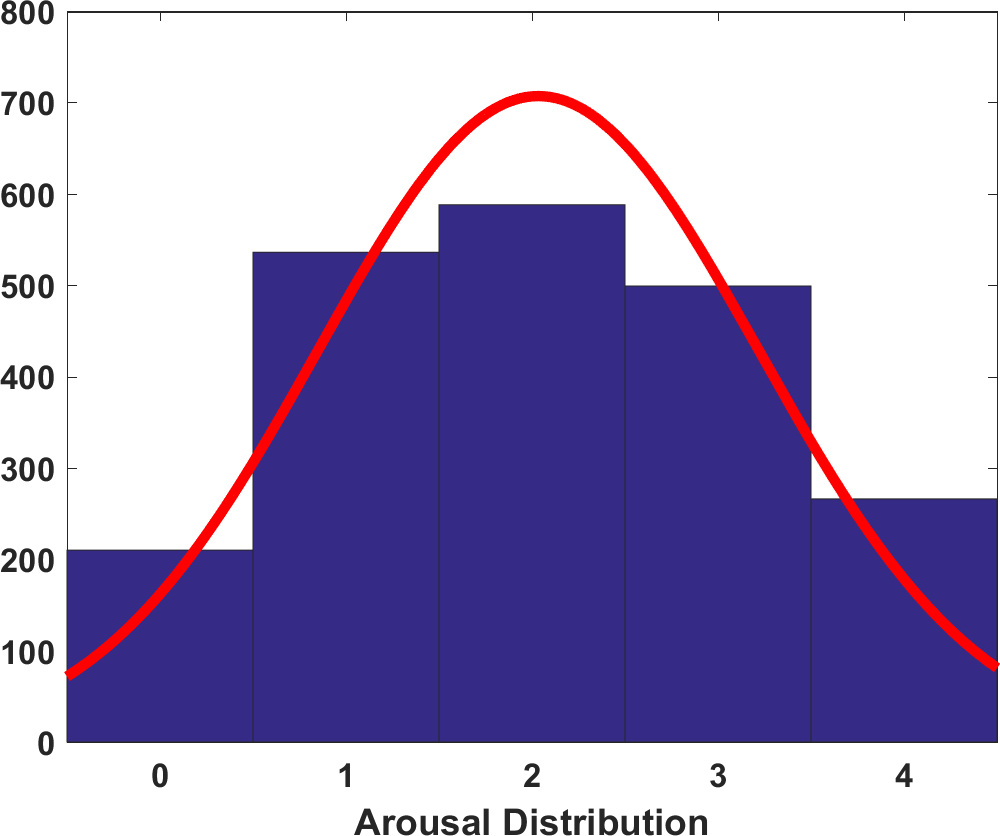}\hfill
\includegraphics[width=0.32\linewidth]{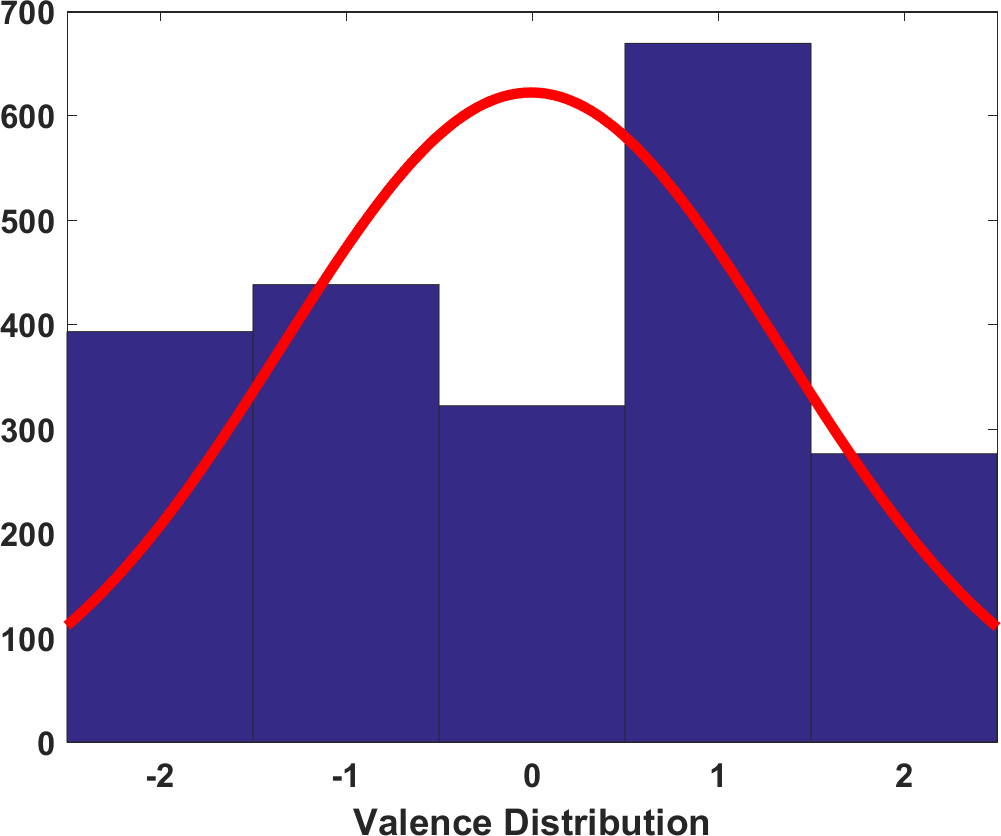}\vspace{-.1cm}
\caption{\label{Annot_dist} (left) Scatter plot of mean asl, val ratings color-coded with expert labels. (middle) Asl and (right) Val rating distribution with Gaussian pdf overlay (view under zoom).}\vspace{-.2cm}
\end{figure}

The \textbf{\textit{circumplex}} emotion model~\cite{Russell1980} defines \textbf{\textit{valence}} as the feeling of \textit{pleasantness}/\textit{unpleasantness} and \textbf{\textit{arousal}} as the \textit{intensity of emotional feeling}. Following this definition, five experts carefully compiled a dataset of 100, roughly 1-minute long commercial ads such that they were uniformly distributed over the arousal--valence plane (Figure~\ref{Annot_dist}). All the 100 ads are publicly available on video hosting websites, and an ad was chosen only if there was consensus among all experts on its valence and arousal labels (categorized as either \textit{high} (H)/\textit{low} (L)). High val ads typically involved product promotions, while low val ads were awareness messages depicting the ill effects of smoking, alcohol and drug abuse, \etc. Expert labels were considered as \textbf{\textit{ground-truth}}, and used for all recognition experiments in this paper. 
 
We then examined if the compiled ads could serve as effective \textit{control} stimuli, \ie, whether they could coherently evoke emotions across viewers. To this end, the 100 ads were independently rated by 23 annotators for val and asl upon familiarizing them with these attributes. All ads were rated on a 5-point scale, which ranged from -2 (\textit{very unpleasant}) to 2 (\textit{very pleasant}) for val and 0 (\textit{calm}) to 4 (\textit{highly aroused}) for asl. Table~\ref{tab:ads_des} presents summary statistics over the four quadrants. In our dataset, low val ads are longer and are perceived as more arousing than high val ads implying that they elicited stronger emotional reactions among viewers.

\begin{table}[t]
\vspace{-.1cm}
\fontsize{7}{7}\selectfont
\renewcommand{\arraystretch}{1.8}
\caption{\label{tab:ads_des} Summary statistics for quadrant-wise ads.}\vspace{-.2cm}
\centering
\begin{tabular}{|c|ccc|} \hline
\textbf{Quadrant} & \textbf{Mean length (s)} & \textbf{Mean asl} & \textbf{Mean val}  \\ \hline \hline
\textbf{H asl, H val} & 48.16 & 2.17 & \ 1.02 \\
\textbf{L asl, H val} & 44.18 & 1.37 & \ 0.91 \\
\textbf{L asl, L val} & 60.24 & 1.76 & -0.76 \\
\textbf{H asl, L val} & 64.16 & 3.01 & -1.16 \\ \hline
\end{tabular}
\vspace{-.3cm}
 \end{table}

To assess whether the compiled ads evoked coherent emotions, we computed agreement among raters in terms of the (a) Krippendorff's $\alpha$, (b) Fleiss $\kappa$ and (c) Cohen's $\kappa$ scores. The $\alpha$ coefficient is applicable when multiple raters rate items ordinally. We obtained $\alpha = 0.62$ (substantial agreement) and $0.36$ (fair agreement) respectively for val and asl, implying that valence impressions were more consistent across raters. On a coarse-grained scale, we computed the Fleiss $\kappa$ agreement among annotators. The Fleiss $\kappa$ statistic (generalization of Cohen's $\kappa$) applies when multiple raters assign categorical values (\textit{high}/\textit{low} in our case) to items. Upon thresholding each rater's asl, val scores by their mean rating to assign \textit{high}/\textit{low} labels for each ad, we observed a Fleiss $\kappa$ of 0.56 (moderate) for valence and 0.27 (fair) for arousal among raters. Computing Fleiss $\kappa$ upon thresholding each rater's scores with respect to the group mean, Fleiss $\kappa$ values of 0.64 (substantial) for val and 0.30 (fair) for asl were noted. Finally, computing Cohen's $\kappa$ agreement between each annotator and groundtruth labels (denoting expert opinion), we obtained a mean Cohen's $\kappa$ of 0.86 (excellent agreement) and 0.68 (substantial agreement) across annotators for val and asl respectively. Overall, these observations convey that (a) greater concordance is noted among novice raters when their opinions are considered collectively rather than individually as subjectivity biases are smoothed out, (b) agreement for val is considerably higher than for asl and (c) the compiled ads evoke consistent affective impressions in the \textit{annotator} and \textit{expert} groups. 

Another desirable property of an affective dataset is the relative independence of the asl and val dimensions~\cite{Russell1980,Barrett99}. To examine asl-val relations for our ad dataset, we (i) examined scatter plots of the annotator ratings, and (ii) computed correlations amongst those ratings. The scatter plot of the mean asl, val annotator ratings, and the distribution of asl and val ratings are presented in Figure~\ref{Annot_dist}. The scatter plot is color-coded based on expert labels, and is interestingly different from the classical `C' shape observed with images~\cite{IAPS}, music videos~\cite{Koelstra} and movie clips~\cite{decaf} attributed to the hypothesis that strong asl evokes strong val ratings. A close examination of the scatter plot reveals that a number of ads are rated as moderate asl, but high/low val. Furthermore, roughly uniform asl and val distributions are observed resulting in Gaussian fits with large variance, especially for val. This is plausible as ads are designed to convey a strong positive or negative message to viewers, while images and movie scenes may convey a relatively neutral emotion. Wilcoxon rank sum tests on ratings revealed significantly different asl ratings for high vs. low asl ads ($p<0.0001$), and distinctive val scores for high vs. low valence ads ($p<0.0001$) consistent with expectation.    

Pearson correlation was computed between the asl and val dimensions with correction for multiple comparisons by limiting the false discovery rate to within 5\%~\cite{benjamini1995controlling}. This procedure revealed a negative and insignificant correlation of 0.17, implying that ad asl and val scores were largely unrelated. Based on the above findings, we claim that our 100 ads constitute a control affective dataset as (i) they induce a fair range of asl and val impressions, which are also found to be largely independent; Different from the `C'-shape characterizing the asl-val relationship for other stimulus types, asl and val ratings are more uniformly distributed for the ad stimuli, and (ii) There is fair-to-substantial concordance among annotators in addition to considerable agreement between novice raters and the ground-truth on affective labels, implying that our ads evoked fairly coherent emotions among viewers.

\subsection{EEG acquisition protocol} As annotators recorded their emotional first impressions on viewing the ads, we acquired their Electroencephalogram (EEG) brain activations via the \textit{Emotiv} wireless headset. The Emotiv device comprises 14 electrodes, and has a sampling rate of 128 Hz. To maximize engagement and minimize fatigue during the rating task, these raters took a break after every 20 ads, and viewed the entire set of 100 ads over five sessions spread over two hours. Each ad was preceded by a 1s fixation cross to orient user attention, and to measure resting state EEG power used for baseline power subtraction. Upon ad viewing, the raters had a maximum of 10 seconds to input their asl and val scores via mouse clicks. Upon experiment completion, the EEG recordings were segmented into \textit{epochs}, with each epoch denoting the time window corresponding to the presentation of the corresponding ad. Upon elimination of corrupted and aborted recordings, we obtained a total of 1738 epochs for 23 viewers. 

\paragraph*{\textbf{Clean vs Raw EEG Data}}
From the recorded 1738 epochs, we manually rejected those epochs which contained head and body movement artifacts. The EEG signal was band-limited between 0.1--45 Hz, and independent component analysis (ICA) was performed to remove artifacts relating to eye movements, eye blinks and muscle movements. This process results in the removal of 212 epochs to leave us with 1526 \textit{clean} epochs. Hereon, \textbf{clean} EEG data will refer to the 1526 preprocessed epochs after visual rejection and ICA, whereas \textbf{raw} EEG data will denote the original 1738 epoch data. We evaluated CNN-based AR performance on both these sets. The following section describes the content and user-centered AR techniques.

\section{Content \& User-centered Analysis}~\label{Data_anal}
\vspace{-.2in}

\subsection{Content-centered Analysis}
For content centered analysis, we extracted and examined audio-visual descriptors from the ads to predict the emotion (in terms of \textit{high}/\textit{low} asl and val) they are likely to evoke. To this end, we employed a deep convolutional neural network (CNN), and the popular handcoded audio-visual descriptors (such as motion activity, audio pitch, \etc.) proposed by Hanjalic and Xu~\cite{Hanjalic2005}. CNNs have recently become very popular for a variety of recognition problems, particularly visual~\cite{alex12} and audio~\cite{Huang2014}, but require vast amounts of labeled training data. As our ad dataset comprised only 100 ads, we fine-tuned the pre-trained \textit{Places205}~\cite{alex12} model via the large-scale and labeled LIRIS-ACCEDE movie dataset~\cite{baveye2015liris}, and employed the fine-tuned model to extract emotional descriptors for our ads. This process is termed as \textit{{domain adaptation}} in machine learning literature.  

To synthesize a deep CNN for ad AR, we employed the pre-trained \textbf{\textit{Places205}} model~\cite{Khosla2013} originally designed for scene understanding. The \textit{Places205} CNN is trained using the {Places-205} dataset comprising 2.5 million images involving 205 scene categories. The {Places-205} dataset contains a wide variety of scenes captured under varying illumination, viewpoint and field of view, and we hypothesized a coherent relationship between scene perspective, lighting and the scene emotion. To find-tune the \textit{Places205} CNN, we employed the labeled \textit{\textbf{LIRIS-ACCEDE}} dataset~\cite{baveye2015liris} which contains asl, val ratings for 9800 $\approx$ 10s long movie snippets. Our ads, on the other hand, are about a minute long with individual ads having lengths ranging from 30--120s.

\subsubsection{FC7 Feature Extraction via CNNs}\label{CNN_anal}
For extracting deep audio-visual features, we input to the \textit{Places205} CNN \textit{key-frame} images for the visual modality, and \textit{spectrograms} for audio. We fine-tuned \textit{Places205} via the LIRIS-ACCEDE~\cite{baveye2015liris} dataset, and employed this model to extract high-level features output by the penultimate fully connected (FC7) CNN layer.

\paragraph*{\textbf{Keyframes as Visual Descriptors}}
From each video in the ad and LIRIS-ACCEDE datasets, we uniformly sampled one \textit{key frame} every three seconds-- this enabled extraction of a continuous video profile for affect prediction. This process generated a total of 1791 key-frames for our 100 ads.

\paragraph*{\textbf{Spectrograms as Audio Descriptors}}

\begin{figure}[t]
\includegraphics[width=0.33\linewidth]{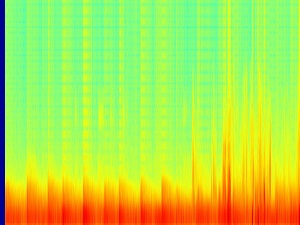}\hfill
\includegraphics[width=0.33\linewidth]{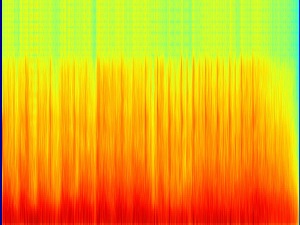}
\includegraphics[width=0.33\linewidth]{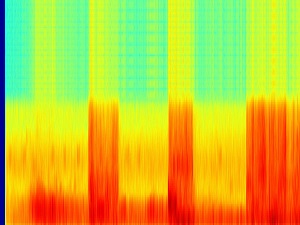}
\centerline{\textbf{LAHV} \hspace{0.11\textwidth} \textbf{HAHV} \hspace{0.11\textwidth} \textbf{HALV}}
\vspace{-.4cm}
\caption{\label{Spect_plot} SGs computed for an exemplar (left) low asl, high val, (middle) high asl, high val and (c) high asl, low val ad. $x$ denotes time (0-10s), while $y$ denotes frequency (Hz). Higher spectral intensities are encoded in yellow and red, and lower intensities are shown in blue and green.}\vspace{-.2cm}
\vspace{-.2cm}
\end{figure}

Spectrograms (SGs) shown in Figure~\ref{Spect_plot} are visual representations of the audio frequency spectrum and have been successfully employed for AR from music and speech~\cite{baveyethesis}. Specifically, transforming the audio content to a spectrogram image allows for audio classification to be treated as a visual recognition problem. We extracted SGs over the 10s long LIRIS-ACCEDE clips and consistently from 10s ad segments. This process generated 610 SGs for our ad dataset. Following~\cite{baveyethesis}, we combined multiple tracks to obtain a single spectrogram (as opposed to two for stereo). Each spectrogram is generated using a 40 ms window short time Fourier transform (STFT), with 20 ms overlap. Larger densities (denoted by red and yellow shades) of high frequencies can be noted in the spectrograms for high asl ads, and these intense scenes are often characterized by high frequency audio (\eg, sudden loud sounds). Conversely, low asl ads tend to retain a sense of continuity in the audio profile, and therefore contain high densities of low frequency sounds.

\paragraph*{\textbf{CNN Training for audio-visual features}}
We used the Caffe~\cite{caffe} deep learning framework for fine-tuning \textit{Places205} with a momentum of 0.9, weight decay of 0.0005, and a base learning rate of 0.0001 reduced by $\frac{1}{10}^{th}$ every 20000 iterations. We totally trained four binary classification networks to recognize high and low asl/val from audio/visual features. To fine-tune \textit{Places205}, we used only the top and bottom 1/3rd LIRIS-ACCEDE videos in terms of asl and val rankings under the assumption that descriptors learned for these extreme-rated clips will effectively model the emotions conveyed by our ads. 4096-dimensional {FC7} layer descriptors extracted from the four networks for our ads were used in the classification experiments. 

\subsubsection{AR with low level audio-visual features}
We benchmark AR performance achieved with CNN features against the handcrafted features proposed by Hanjalic and Xu~\cite{Hanjalic2005}. Even after a decade, their work remains one of the most popular AR baselines as seen from recent works such as~\cite{Koelstra,decaf}. In~\cite{Hanjalic2005}, asl and val are modeled via low-level descriptors describing motion activity, colorfulness, shot change frequency, voice pitch and sound energy in the scene. These predictors are intuitive and interpretable, and are used to estimate time-continuous asl and val levels in~\cite{Hanjalic2005}. Table~\ref{tab:exp_det} summarizes audiovisual features for content-centric AR, and the proportion of positive class samples for val and asl available with audio and video modalities. We attempt asl/val classification at the keyframe/spectrogram level, and class probabilities are aggregated to obtain ad-level scores for the application presented in Sec.~\ref{user_study}.

\subsection{User-centered analysis}
The 1738 epochs obtained from the EEG acquisition process were used for user-centered analysis. However, these epochs were of different lengths as ad durations were variable. To maintain dimensional consistency, we performed user-centric AR experiments with (a) the \textit{first} 3667 samples ($\approx 30s$ of EEG data), (b) the \textit{last} 3667 samples and (c) the \textit{last} 1280 samples (10s of EEG data) from each epoch(consistent with content-centered analysis, and to examine temporal effects on AR). Each epoch sample comprised data from 14 EEG channels, and all epoch samples were input to a classifier upon vectorization. In addition to conventional classifiers, we also used a deep neural network to classify EEG epochs whose architecture is described below.

\subsubsection{EEG Feature Extraction for CNN Training}\label{EEG-CNN}
As we had a relatively small number of epochs (1738) with very high dimensionality (14 channels $\times$ 3667 time points = {51338} dimensional vector), a CNN trained on this data is highly susceptible to overfitting. To alleviate overfitting, we applied Principal Component Analysis (PCA) on the vectorized epochs to reduce dimensionality. PCA has been successfully employed for CNN-based EEG classification recently~\cite{jirayucharoensak2014eeg, siuly2016injecting}, and a number of works have stressed the need for PCA-based pre-processing for robust EEG signal representation prior to neural network training \cite{kavasidis2017brain2image,spampinato2017deep,stober2014using,stober2015deep,stober2017learning}. Specifically, \cite{jirayucharoensak2014eeg} discusses PCA effectiveness for deriving a good EEG input representation for CNNs.

\paragraph*{\textbf{CNN Training for EEG features}}
The dimensionality-reduced EEG features (preserving 90\% data variance) were then passed to a CNN for val, asl recognition. We used a CNN architecture employed for time-series sensor data classification \cite{rad2018deep} and implemented with the Keras \cite{chollet2015keras} library. The network is three layers deep with two 1-D convolutional layers followed by a fully connected layer. Training was performed with 64 $1 \times 3$ filters in the 1-D convolutional layers and 128 nodes in the fully connected layer. We set a momentum factor of 0.9, weight decay of 0.0005 and a base learning rate of 0.0001. A dropout level of 0.5 was used to prevent overfitting. The model was trained for a maximum of 100 epochs, and early stopping was forced in case the validation loss increased over five successive training iterations. For both content and user-centric analysis, 80\% of the compiled dataset was used for training and the remaining 20\% for testing with the process repeated 10 times (10 $\times$ 5-fold cross validation).  

\begin{table}[t]
\fontsize{6.5}{6.5}\selectfont
\renewcommand{\tabcolsep}{3.8pt}
\caption{Extracted features for content-centric AR. +ve class proportions (as \%) for val/asl in the audio and visual modalities are specified.} \label{tab:exp_det} \vspace{-.5cm}
\begin{center}                                                         
\begin{tabular}{@{}|c|ccc|@{}} 

\hline
\textbf{Attribute} & \multicolumn{3}{c|}{\textbf{Valence/Arousal}} \\ 
\textbf{Descriptors} &{\textbf{Audio}}& {\textbf{Video}}&{\textbf{aud+vid (A+V)}}\\ \hline\hline
 
 {\textbf{CNN}} & 4096D FC7 features  & 4096D FC7 features & 8192D FC7 features \\
 \textbf{Features} &  from 10s SGs.    & extracted from keyframes &from SGs + keyframes \\ 
        & & sampled every 3 seconds.  & over 10s intervals. \\ \hline

\textbf{Hanjalic~\cite{Hanjalic2005}} & Per-second sound  & Per-second shot change & Concatenation of \\ 
\textbf{Features} &  energy and pitch  & frequency and motion & audio-visual features. \\
 & statistics~\cite{Hanjalic2005}. & statistics~\cite{Hanjalic2005}. & ~ \\ \hline
\textbf{+ve class}  & {43.8}/{51.9} & {43.4}/{51.6} & {43.8}/{51.9}\\ 
\textbf{prop (\%)}  & & & \\ \hline
\end{tabular}
\end{center}
\vspace{-.5cm}
\end{table}

\begin{figure*}[t]
\centerline{\includegraphics[width=0.33\linewidth,height=3.5cm]{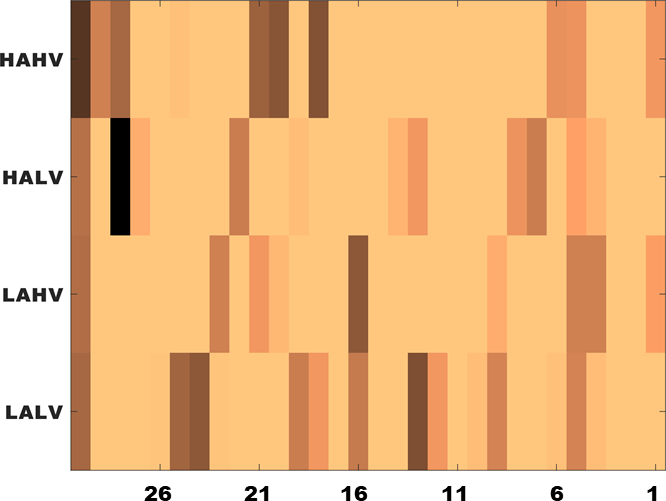}\hspace{0.05cm}
\includegraphics[width=0.33\linewidth,height=3.5cm]{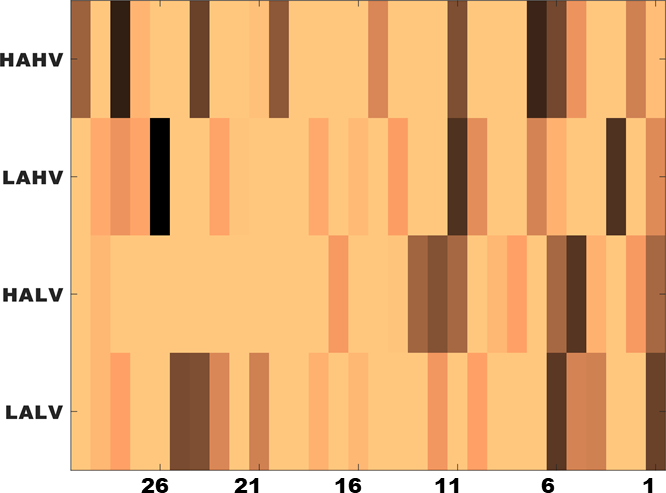}\hspace{0.05cm}
\includegraphics[width=0.33\linewidth,height=3.5cm]{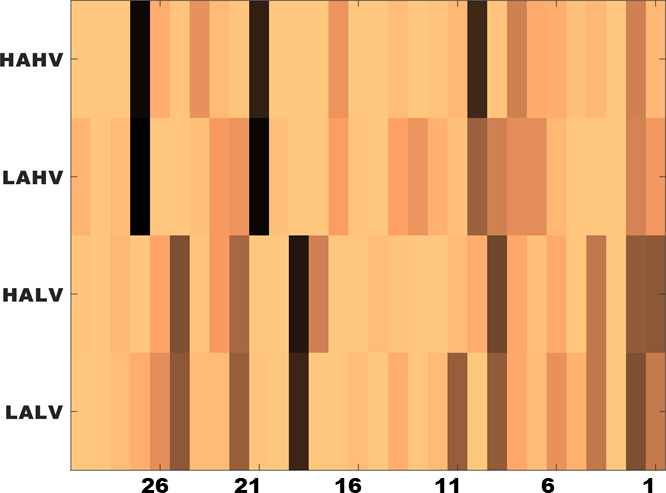}}
\centerline{\textbf{Shot Frequency}\hspace{0.21\linewidth}\textbf{Pitch Amplitude}\hspace{0.23\linewidth}\textbf{Motion Activity}} \vspace{-.2cm} 
\caption{\label{MTL_ex} Learned MTL weights for  the four quadrants (tasks) when fed with the specified low-level features computed over the final 30s of 100 ads.}
\vspace{-.3cm}
\end{figure*}



\section{Experiments and Results}~\label{ER}
We first describe classifiers and settings employed for binary content and user-centric AR, where the objective is to assign a binary (H/L) label for asl and val evoked by each ad, using the extracted fc7/low-level audiovisual/EEG  features. Ad labels are provided by experts, whose opinions greatly agreed with naive users (Sec~\ref{DD}). 

\begin{table*}[htbp]
\fontsize{8}{8}\selectfont
\renewcommand{\arraystretch}{1.3}
\centering
\caption{Ad AR from content analysis. F1 scores are presented in the form $\mu \pm \sigma$. } \label{tab:ccap} \vspace{-.3cm}
\begin{tabular}{|c|ccc|ccc|}
  \hline
 
	\multicolumn{1}{|c|}{\textbf{Method}} & \multicolumn{3}{c|}{\textbf{Valence}} & \multicolumn{3}{c|}{\textbf{Arousal}} \\ \hline 
	\multicolumn{1}{|c|}{~} & {\textbf{F1 (all)}} & {\textbf{F1 (L30)}} & {\textbf{F1 (L10)}}  & {\textbf{F1 (all)}} & {\textbf{F1 (L30)}} & {\textbf{F1 (L10)}}\\ \hline

	
	\textbf{Audio FC7 + LDA}  & 0.61$\pm$0.04 & 0.62$\pm$0.10 & 0.55$\pm$0.18 & 0.65$\pm$0.04 & 0.59$\pm$0.10 & {0.53$\pm$0.19}\\
	\textbf{Audio FC7 + LSVM} & 0.60$\pm$0.04 & 0.60$\pm$0.09 & 0.55$\pm$0.19 & 0.63$\pm$0.04 & 0.57$\pm$0.09 & 0.50$\pm$0.18\\
	\textbf{Audio FC7 + RSVM} & {0.64$\pm$0.04} & \textbf{0.66$\pm$0.08} & {0.62$\pm$0.17} & \textbf{0.68$\pm$0.04} & {0.60$\pm$0.10} & {0.53$\pm$0.19}\\ \hline

	\textbf{Video FC7 + LDA} & 0.69$\pm$0.02 & 0.79$\pm$0.08 & {0.77$\pm$0.13} & 0.63$\pm$0.03 & 0.58$\pm$0.10 & 0.57$\pm$0.18\\
	\textbf{Video FC7 + LSVM}& 0.69$\pm$0.02 & 0.74$\pm$0.08 & 0.70$\pm$0.15 & 0.62$\pm$0.02 & 0.57$\pm$0.09 & 0.52$\pm$0.17\\
	\textbf{Video FC7 + RSVM}& {0.72$\pm$0.02} & \textbf{0.79$\pm$0.07} & 0.74$\pm$0.15 & \textbf{0.67$\pm$0.02} & {0.62$\pm$0.10} & {0.58$\pm$0.19}\\ \hline
    
    	\textbf{Audio FC7 + MTL} & 0.85$\pm$0.02 & 0.83$\pm$0.10 & 0.78$\pm$0.20& 0.78$\pm$0.03 & 0.62$\pm$0.14 & 0.45$\pm$0.16\\ 
	\textbf{Video FC7 + MTL} &\textbf{0.96$\pm$0.01} & 0.94$\pm$0.07 & 0.82$\pm$0.25 & \textbf{0.94$\pm$0.01} & 0.87$\pm$0.12 & 0.63$\pm$0.29 \\ \hline    
    
	\textbf{{A+V FC7 + LDA}} &  0.70$\pm$0.04 & 0.66$\pm$0.08 & 0.49$\pm$0.18 & 0.60$\pm$0.04 & 0.52$\pm$0.10 & {0.51$\pm$0.18}\\
	\textbf{A+V FC7 + LSVM}  &  0.71$\pm$0.04 & 0.66$\pm$0.07 & 0.49$\pm$0.19 & 0.56$\pm$0.04 & 0.49$\pm$0.10 & 0.47$\pm$0.19\\
	\textbf{A+V FC7 + RSVM}  &  \textbf{0.75$\pm$0.04} & {0.70$\pm$0.07} & {0.55$\pm$0.17} & \textbf{0.63$\pm$0.04} & {0.56$\pm$0.11} & 0.49$\pm$0.19\\ \hline
	
\textbf{{A+V Han + LDA}} & 0.59$\pm$0.09 &{0.63$\pm$0.08} & {0.64$\pm$0.12} & {0.54$\pm$0.09} & 0.50$\pm$0.10 & {0.58$\pm$0.08}\\
	\textbf{A+V Han + LSVM}  & {0.62$\pm$0.09} & {0.62$\pm$0.10} & {0.65$\pm$0.11} & 0.55$\pm$0.10 & {0.51$\pm$0.11} & 0.57$\pm$0.09\\
	\textbf{A+V Han + RSVM}  & \textbf{0.65$\pm$0.09} & {0.62$\pm$0.11} & 0.62$\pm$0.12 & \textbf{0.59$\pm$0.12} & {0.58$\pm$0.11} & 0.56$\pm$0.10\\ \hline
		
	\textbf{{A+V FC7 LDA DF}} & 0.60$\pm$0.04  &  {0.66$\pm$0.04}  &  {0.70$\pm$0.19} & 0.59$\pm$0.02 & 0.60$\pm$0.07 & {0.57$\pm$0.15} \\
	\textbf{A+V FC7 LSVM DF}  & {0.65$\pm$0.02}  &  {0.66$\pm$0.04}  &  0.65$\pm$0.08 & {0.60$\pm$0.04}  & {0.63$\pm$0.10} & {0.53$\pm$0.13} \\
	\textbf{A+V FC7 RSVM DF}  & \textbf{{0.72$\pm$0.04}}  &  {0.70$\pm$0.04}  &  {0.70$\pm$0.12} & {0.69$\pm$0.06} & \textbf{0.75$\pm$0.07} & {0.70$\pm$0.07}\\ \hline
	
\textbf{{A+V Han LDA DF}}   & 0.58$\pm$0.09 & 0.58$\pm$0.09 & \textbf{0.61$\pm$0.09} & {0.59$\pm$0.06} & {0.59$\pm$0.07} & {0.61$\pm$0.08}\\
	\textbf{A+V Han LSVM DF}  & 0.59$\pm$0.10 & 0.59$\pm$0.09 & 0.60$\pm$0.10 & {\textbf{0.61$\pm$0.05}} & {0.61$\pm$0.08} & {0.60$\pm$0.09}\\
	\textbf{A+V Han RSVM DF}  & 0.60$\pm$0.08 & 0.56$\pm$0.10 & {0.58$\pm$0.09} & {0.58$\pm$0.09} & {0.56$\pm$0.06} & {0.58$\pm$0.09}\\ \hline
    
	\textbf{A+V FC7 + MTL}  & \textbf{{0.89$\pm$0.03}} & {0.88$\pm$0.11} & {0.77$\pm$0.26} & \textbf{{0.87$\pm$0.03}} & {0.68$\pm$0.17} & {0.46$\pm$0.20} \\
	%
	
	\textbf{A+V Han + MTL}  & 0.77$\pm$0.04 & 0.79$\pm$0.07 & 0.74$\pm$0.15 & 0.78$\pm$0.04 & 0.73$\pm$0.11 & 0.58$\pm$0.22\\ \hline
	
  \hline
\end{tabular}
\vspace{0.1cm}
%
%
\fontsize{8}{8}\selectfont
\renewcommand{\arraystretch}{1.3}
\centering
\caption{Ad AR from EEG analysis. F1 scores are presented in the form $\mu \pm \sigma$. } \label{tab:uscap} \vspace{-.2cm}
\begin{tabular}{|c|ccc|ccc|}
  \hline
	\multicolumn{1}{|c|}{\textbf{Method}} & \multicolumn{3}{c|}{\textbf{Valence}} & \multicolumn{3}{c|}{\textbf{Arousal}} \\ \hline 
	\multicolumn{1}{|c|}{~} & {\textbf{F1 (F30)}} & {\textbf{F1 (L30)}} & {\textbf{F1 (L10)}}  & {\textbf{F1 (F30)}} & {\textbf{F1 (L30)}} & {\textbf{F1 (L10)}}\\ \hline

	
	
    \textbf{Raw EEG $+$ LDA } & 0.79 $\pm$ 0.02 & 0.78 $\pm$ 0.02 & 0.76 $\pm$ 0.03 & 0.76 $\pm$ 0.02 & 0.76 $\pm$ 0.02 & 0.72 $\pm$ 0.04\\
		\textbf{Raw EEG $+$ LSVM } & 0.78 $\pm$ 0.03 & 0.77 $\pm$ 0.04 & 0.77 $\pm$ 0.05 & 0.75 $\pm$ 0.03 & 0.74 $\pm$ 0.02 & 0.70 $\pm$ 0.04\\
		\textbf{Raw EEG $+$ RSVM } & \textbf{{{0.80 $\pm$ 0.03}}} & {{0.79 $\pm$ 0.03}} & {{{0.79 $\pm$ 0.03}}} & \textbf{{0.77 $\pm$ 0.03}} & {{{0.77 $\pm$ 0.04}}} & {{{0.74 $\pm$ 0.04}}}\\ \hline
		
	\textbf{Clean EEG $+$ LDA} & 0.79 $\pm$ 0.03 & 0.79 $\pm$ 0.03 & 0.77 $\pm$ 0.03 & 0.76 $\pm$ 0.03 & 0.75 $\pm$ 0.03 & 0.71 $\pm$ 0.04\\	
	\textbf{Clean EEG $+$ LSVM } & 0.77 $\pm$ 0.03 & 0.76 $\pm$ 0.04 & 0.77 $\pm$ 0.05 & 0.74 $\pm$ 0.03 & 0.73 $\pm$ 0.02 & 0.69 $\pm$ 0.04\\
	\textbf{Clean EEG $+$ RSVM } & \textbf{{{0.82 $\pm$ 0.03}}} & \textbf{{0.82 $\pm$ 0.03}} & {{{0.81 $\pm$ 0.03}}} & \textbf{{0.78 $\pm$ 0.02}} & {{{0.77 $\pm$ 0.03}}} & {{{0.75 $\pm$ 0.04}}}\\ \hline

		\textbf{Raw EEG $+$ CNN} & {{0.85 $\pm$ 0.03}} & {0.85 $\pm$ 0.03} & {{0.83 $\pm$ 0.03}} & {0.84 $\pm$ 0.02} & {{0.82 $\pm$ 0.03}} & {{0.79 $\pm$ 0.04}}\\
    \textbf{Clean EEG $+$ CNN} & \textbf{{0.89 $\pm$ 0.05}} & {0.88 $\pm$ 0.04} & {{0.88 $\pm$ 0.05}} & \textbf{0.87 $\pm$ 0.03} & {{0.85 $\pm$ 0.04}} & {{0.80 $\pm$ 0.06}}\\ \hline
		
     \textbf{Raw EEG $+$ MTL} & {{0.92 $\pm$ 0.01}} & {0.91 $\pm$ 0.01} & {{0.90 $\pm$ 0.01}} & {0.90 $\pm$ 0.02} & {{0.87 $\pm$ 0.04}} & {{0.85 $\pm$ 0.05}}\\
    \textbf{Clean EEG $+$ MTL} & \textbf{{0.97 $\pm$ 0.01}} & \textbf{0.97 $\pm$ 0.01} & {{0.93 $\pm$ 0.03}} & \textbf{0.96 $\pm$ 0.01} & {{0.94 $\pm$ 0.02}} & {{0.90 $\pm$ 0.04}}\\
   
    \hline
\end{tabular}
\vspace{0.2cm}
%
%
\fontsize{8}{8}\selectfont
\renewcommand{\arraystretch}{1.3}
\centering
\caption{Probablistic fusion of audiovisual \& EEG classifier outputs. F1 scores are presented in the form $\mu \pm \sigma$. } \label{tab:uscapfus} \vspace{-.2cm}
\begin{tabular}{|c|ccc|ccc|}
  \hline
	\multicolumn{1}{|c|}{\textbf{Method}} & \multicolumn{3}{c|}{\textbf{Valence}} & \multicolumn{3}{c|}{\textbf{Arousal}} \\ \hline 
	\multicolumn{1}{|c|}{~} & {\textbf{F1 (F30)}} & {\textbf{F1 (L30)}} & {\textbf{F1 (L10)}}  & {\textbf{F1 (F30)}} & {\textbf{F1 (L30)}} & {\textbf{F1 (L10)}}\\ \hline
	\textbf{(Raw EEG + RSVM) + (A+V fc7 RSVM) DF} & 0.85 $\pm$ 0.03 & 0.84 $\pm$ 0.03 & 0.84 $\pm$ 0.03 & 0.84 $\pm$ 0.03 & 0.83 $\pm$ 0.03 & 0.80 $\pm$ 0.04\\
    \textbf{(Raw EEG + CNN) + (Audiovisual fc7 RSVM) DF} & 0.87 $\pm$ 0.03 & 0.87 $\pm$ 0.03 & 0.86 $\pm$ 0.02 & 0.86 $\pm$ 0.01 & 0.85 $\pm$ 0.03 & 0.83 $\pm$ 0.04\\
	\textbf{(Clean EEG + RSVM) + (A+V fc7 RSVM) DF} & 0.86 $\pm$ 0.03 & 0.85 $\pm$ 0.03 & 0.86 $\pm$ 0.03 & 0.85 $\pm$ 0.02 & 0.83 $\pm$ 0.04 & 0.82 $\pm$ 0.04\\
    \textbf{(Clean EEG + CNN) + (A+V fc7 RSVM) DF} & \textbf{0.91 $\pm$ 0.03} & 0.89 $\pm$ 0.03 & 0.88 $\pm$ 0.02 & \textbf{0.88 $\pm$ 0.02} & 0.87 $\pm$ 0.02 & 0.84 $\pm$ 0.04\\
    \hline
\end{tabular}
\vspace{-.5cm}
\end{table*}

\paragraph*{\textbf{Classifiers}} We considered both shallow and deep classifiers for content and user-centered AR. Among shallow classifiers, we employed linear discriminant analysis (LDA), linear SVM (LSVM) and radial basis function SVM (RSVM). LDA and LSVM partition training data via a separating hyperplane, while RSVM transforms input data onto a high-dimensional feature space where the positive and negative class samples can be linearly separated. FC7 features learned from audiovisual descriptors (Sections~\ref{CNN_anal} and~\ref{EEG-CNN}) were input to shallow classifiers for content-centered analysis, while EEG descriptors were fed to both shallow classifiers and the three layer CNN for user-centered AR.

In addition to the above \textit{single-task learning} methods which do not exploit the underlying structure of the input data, we also explored the use of \textit{multi-task learning} (MTL) for AR. When posed with the learning of multiple \textit{related} tasks, MTL seeks to jointly learn a set of task-specific classifiers on modeling task relationships, which is highly beneficial when learning with few examples. Among the MTL methods available as part of the MALSAR package~\cite{zhou2012mutal}, we employed the sparse graph-regularized MTL (SR-MTL) where \textit{a-priori} knowledge regarding task-relatedness is modeled in the form of a graph $R$. Given tasks $t = 1...T$, with $X_t$ denoting training data for task $t$ and $Y_t$ their labels, SR-MTL jointly learns a weight matrix $W = [W_1 \ldots W_T]$ such that the objective function $\sum_{t=1}^{T} \Vert W_t^T X_t -Y_t\Vert_F^2+\alpha \Vert WR \Vert^2_{F}+ \beta \Vert W \Vert_1+\gamma \Vert W \Vert^2_F$ is minimized. Here, $\alpha, \beta, \gamma$ are regularization parameters, while  $\Vert.\Vert_F$ and $\Vert.\Vert_1$ denote matrix Frobenius ($\ell_2$) and $\ell_1$-norms.

MTL is particularly suited for dimensional AR, and one can expect similarities in terms of audio-visual content among high val or high asl ads. We exploit underlying similarities by modeling each asl-val quadrant as a task (\ie, all H asl, H val ads will have identical task labels). Also, quadrants with same asl/val labels are deemed as \textit{related} tasks, while those with dissimilar labels are considered \textit{unrelated}. Task relatedness is then modeled via edge weights 
$\gamma_{ij}$ for the graph $R$, \ie, $\gamma_{ij} = 1$ for related tasks, and $\gamma_{ij} = 0$ for unrelated tasks, where ${i,j} \in 1 \ldots T, i \neq j$.

The graph $R$ then guides the learning of $W_t$'s as shown in the three examples in Fig.\ref{MTL_ex}, where SR-MTL is fed with the specified features computed over the final 30 s of all ads. Darker shades denote salient MTL weights. Shot change frequency is found to be a key predictor of asl~\cite{Hanjalic2005}, and one can notice salient weights for H asl, H val ads in particular. The attributable reason is that our H asl H val ads involve frequent shot changes to maintain emotional intensity, while the mood of our H asl, L val ads is strongly influenced by semantics ({depicting topics like drug and alcohol abuse, and overspeeding}). Likewise, pitch amplitude is a key val predictor, and salient weights can be consistently seen over the 30s temporal window for HV ads. Finally, more salient weights for H val ads with motion activity reveals that our positive val ads involve accentuated motion.   

For content-centric AR, apart from unimodal (audio (A) or visual (V)) fc7 features, we also employed \textit{feature fusion} (A+V entries in Table~\ref{tab:ccap}). Probabilistic \textit{decision fusion} of the unimodal classifier outputs was attempted with audiovisual features (A$+$V DF entries in Table~\ref{tab:ccap}), and with audiovisual and EEG (Audiovisual $+$ EEG DF in Table~\ref{tab:uscapfus}) features. Audiovisual feature fusion (A$+$V) involved concatenation of fc7 A and V features over 10s windows (see Table~\ref{tab:exp_det}), while the $W_{est}$ technique~\cite{koelstra2012fusion} was employed for decision fusion (DF). In DF, the test label is assigned the index $j$, $j \in$ \{H(1) ,L(0)\}, corresponding to maximum class probability $P_j = \sum_{i=1}^2 \alpha_i^*t_ip_i$,  where $i$ denotes the constituent modalities, $p_i$'s denote classifier posteriors and $\{\alpha_i^*\}$ are the optimal weights maximizing test F1-score determined via a 2D grid search. If $F_i$ denotes the training F1-score for the $i^{th}$ modality, then $t_i = \alpha_i F_i/\sum_{i=1}^2 \alpha_i F_i$ for given $\alpha_i$. 

\paragraph*{\textbf{Metrics and Experimental Settings}} We used the F1-score (F1) defined as the harmonic mean of precision and recall for evaluation. F1-score is appropriate for our setting due to the imbalance in the +ve and -ve class proportions. We compare our audiovisual fc7 and EEG features against the baseline features of Hanjalic and Xu~\cite{Hanjalic2005}. These hand-crafted features are interpretable, and employed to estimate time-continuous asl and val levels. As the Hanjalic (Han) algorithm~\cite{Hanjalic2005} inherently uses audiovisual features to model asl and val, we only consider (feature and decision) fusion performance in this case. User-centered AR uses only PCA-applied EEG features (Sec.~\ref{EEG-CNN}).
AR results obtained over ten repetitions of 5-fold cross validation (CV) (50 runs) are presented. CV is used to address the \textit{overfitting} problem on small datasets, and optimal SVM parameters are determined via an inner five-fold CV on the training set. To examine the temporal variance in AR performance, we present F1-scores obtained over (a) all ad frames (`All'), (b) last 30s (L30) and (c) last 10s (L10) for \textit{content-centered} AR. Similarly, AR results are presented for (a) first 30s (F30), (b) last 30s (L30) and (c) last 10s (L10) for \textit{user-centered} AR. These settings were chosen as EEG sampling rate is higher than for audio/video.  

\subsection{Results Overview}
Tables~\ref{tab:ccap} and \ref{tab:uscap} respectively present content-centric and user-centric AR results for the various settings described above, whereas Table~\ref{tab:uscapfus} presents results on fusing the audiovisual and EEG-based classifier outputs. The highest F1 score achieved for a given modality across all classifiers and temporal settings is denoted in bold. \\
\indent{\textbf{Content-centric analysis:}} Focusing on unimodal descriptors in Table~\ref{tab:ccap}, we note that video fc7 features predict val (peak F1 = 0.79) considerably better than asl, while audio fc7 features encode asl (peak F1 = 0.68) slightly better than val (peak F1 = 0.66). Also, much superior AR is achieved with MTL (peak F1 = 0.96 for val, 0.94 for asl) as compared to single task classifiers. With single and multi-task classifiers, consistently higher F1-scores are noted with video fc7 features, implying that better emotion predictors are learned from the raw video data as compared to spectrograms. 

Concerning multimodal methods, we firstly note that multimodal approaches achieve comparable or better F1 scores as compared to unimodal ones. For val, the best fusion performance (F1 = 0.75 with feature fusion and RSVM classifier) is superior compared to audio-based (F1 = 0.66), but inferior compared to video-based (F1 = 0.79) recognition. Contrastingly for asl, fusion F1-score (0.75 with DF) considerably outperforms unimodal methods (0.68 with audio, and 0.67 with video). 
We first examine feature-fusion approaches. Comparing A$+$V {{fc7 vs Han}} features, fc7 descriptors clearly outperform Han features with both single and multi-task methods. Performance difference is prominent for val (F1 = 0.75 with fc7 vs 0.65 with Han), while comparable recognition is achieved with either feature for asl (F1 of 0.63 with fc7 vs 0.59 with Han).   

Examining AR performance with decision fusion (DF) methods, DF (F1 = 0.75) substantially outperforms feature fusion (F1 = 0.59) for asl recognition, while underpeforming for val (F1 = 0.72 with DF and 0.75 with feature fusion). Among classifiers, RSVM produces the best F1-scores for both asl and val among {{single-task classifiers}} with both unimodal and multimodal features. This indicates that the fc7 audiovisual features may not be easily linearly separable in the respective feature spaces. Nevertheless, the linear {{MTL}} model beats all single-task methods with both fc7 and Han features. MTL F1-scores in the {A+V FC7 + MTL} condition are considerably higher than single-task multimodal F1-scores, and the trend repeats with unimodal features as well.  These observations suggest that learning underlying feature similarities among ads with \textit{similar attributes} enables better separability of H and L asl/val data.

\paragraph*{\textbf{EEG-based AR}} From user-centered analysis, we mainly examine whether (a) better AR is achievable by examining user responses vis-\~a-vis mining the audiovisual ad content; (b) the three-layer CNN (Section~\ref{EEG-CNN}) better encodes emotional attributes as compared to shallow classifiers, and (c) whether the considered CNN architecture could achieve similar AR performance with \textit{clean} vs \textit{noisy} EEG signals, as CNNs are adept at learning target encodings from disparate data.  

Observing Table~\ref{tab:uscap}, we firstly observe that EEG-based AR results are generally superior to \textit{content-centric} results. The best EEG-based val and asl F1-scores are considerably higher than the best {content-centered} unimodal results. As with audiovisual features, EEG achieves better val recognition than asl different from findings reported in~\cite{subramanian2016ascertain,Koelstra,decaf}. In this regard, we observe that positive val is found to correlate with increased activity in the frontal lobes~\cite{oude2006eeg}, and the \textit{Emotiv} device efficiently captures frontal lobe activity despite its limited spatial resolution. 

Among the shallow classifiers considered with EEG data, RSVM again performs best. The three-layer CNN however outperforms shallow classifiers by far. Also, while very comparable results are achieved with the raw (noisy) and cleaned EEG data for val and asl employing shallow classifiers, larger performance differences are noted with the CNN and MTL methods. This implies that that while all approaches are able to discriminate the noisy EEG features, CNN and MTL are able to discriminate better from the cleaned data. As with audiovisual descriptors, highest F1-scores with EEG features (close to ceiling performance for both val and asl) are also obtained with MTL, reinforcing its utility for emotion recognition.\\
\indent{\textbf{General Observations:}} Relatively small $\sigma$ values are observed in the `All' condition with both audiovisual and EEG-based CNN features for 
the five-fold CV procedure in Tables~\ref{tab:ccap} and~\ref{tab:uscap}. These trends suggest that the corresponding classification models do not {{overfit}}. Examining {{temporal windows}} considered for audiovisual AR, significantly higher $\sigma$'s are nevertheless noted with Han features as well as with the L30 and L10 temporal segments, conveying that the corresponding models do not generalize well. Higher $\sigma$'s observed for the L30 and L10 conditions reveal considerable variance in AR performance on the terminal ad frames. Contrastingly, very similar $\sigma$'s are noted for the different temporal windows considered with EEG data in Table~\ref{tab:uscap}. 

Interestingly in Table~\ref{tab:ccap}, one can note a considerable decrease in asl F1 scores for the L30 and L10 conditions with audio and visual features, while val F1-scores are similar to the 'All' condition. Also, a sharp degradation in MTL performance is noted in the L30 and L10 conditions.  Corresponding inferences are tabulated as follows. (1) Greater differences between ads towards endings are characterized by the large F1 variance in the L30 and L10 conditions with unimodal and multimodal features; conversely, similar AR performance is noted with EEG features for the different temporal segments. This implies that while the audiovisual information conveying ad emotion may significantly vary over time, human viewers typically tend to grasp the conveyed emotion rather instantaneously; (2) Fusion models synthesized with Han features are most prone to overfitting, given the generally larger $\sigma$ values seen with respect to other models. (3) Lower asl F1 scores in the L30 and L10 conditions highlight the limitation of using a \textit{single} asl/val label (as opposed to dynamic labeling) over time. Generally lower F1-scores achieved for asl with all methods in Table~\ref{tab:ccap} suggests that asl is more difficult to characterize than val (this could possibly explain the lower agreement for asl in Section~\ref{DD}), while coherency between val features and labels remains sustained over time.

\paragraph*{\textbf{Fusion of Content and User-Centric Modalities}}
Given the difference in AR performance observed on mining the content and user-centered descriptors (especially with respect to variance across temporal segments), one could possibly conclude that the audiovisual and EEG modalities encode complementary information. Therefore, we examined if probabilistic fusion of the content (A$+$V fc7) and EEG-based classifier outputs resulted in better asl/val recognition. Corresponding results are tabulated in Table~\ref{tab:uscapfus}.  

Comparing Table~\ref{tab:uscapfus} against Tables~\ref{tab:ccap} and~\ref{tab:uscap} clearly reveals that fusing complementary information is beneficial. Fusion-based asl and val F1-scores are consistently better than individual counterparts, and more superior when shallow classifiers are employed to perform individual predictions (rows 1 and 3). These findings reveal the potential for fusion of {content} and {user-centric} cues as in~\cite{nusef2010,cntxtarxiv,retar2013}. 

\section{Computational Advertising - User Study}~\label{user_study}
Presented results clearly reveal that the compiled fc7 audiovisual and EEG desciptors outperform the baseline Han features for ad AR. We now demonstrate that improved AR positively impacts a computational advertising application-- specifically, we show that better AR facilitates \textit{optimized} insertion of ads onto streamed (\eg, YouTube) video. We characterize {optimized} ad insertion in terms of twin (possibly conflicting) objectives: (1) maximizing ad impact (measured in terms of ad memorability), and (2) minimially disrupting (or ideally enhancing) the viewing experience.  

The research question that we seek to study here is 
\textit{Whether better affect estimation, as achieved by the CNN frameworks harnessing audiovisual and EEG descriptors, leads to optimal insertion of ads at appropriate scene transition points in a video sequence?} A principled methodology to insert ads in video is proposed by the CAVVA algorithm~\cite{cavva}. CAVVA is a genetic algorithm-based  optimization framework for inserting ads onto streamed video. On top of low-level feature based contextual matching as proposed by frameworks such as VideoSense~\cite{videosense}, CAVVA models affective relevance between scenes in a video sequence and ads in an inventory to determine the (a) \textit{suitable} ads to insert, and (b) the \textit{best} temporal positions in the video sequence where the chosen ads should be inserted. 

Based on consumer psychology insights, CAVVA proposes ad insertion rules that seek to strike a balance between (a) maximizing ad impact in terms of brand memorability (\ie, maximizing ad recall), and (b) minimally disrupting (or even enhancing) viewer engagement and experience. To examine the above research question, we performed a study with 18 users to compare ad recall and subjective quality of advertising schedules generated with affective scores estimated via (a) the content-centric audiovisual CNN model, (b) the user-centric EEG CNN model and (c) first impression ratings provided by experts. Details of the (i) ad and video datasets employed, (ii) employed ad insertion strategies and (iii) user study and associated results are as follows.

\subsection{Ad and Video Datasets}\label{US-DS} 

For the user study, we used 28 ads (from the original 100), and three program videos. The ads were equally distributed among the four quadrants of the asl-val plane based on expert labels. The program videos were scenes from a television sitcom (\textit{friends}) and two movies (\textit{ipoh} and \textit{coh}), which predominantly comprised social themes and situations capable of invoking high-to-low valence and moderate arousal (see Table~\ref{tab:progdetails} for statistics). Each program video comprised eight scenes implying that there were seven possible ad-insertion points corresponding to scene transitions. The average scene length in the program videos was 118 seconds.

\begin{table}[t]
\fontsize{8}{8}\selectfont
\renewcommand{\arraystretch}{1.3}
\centering
\caption{Summary of program video statistics.}\vspace{-.3cm}
\begin{tabular}{|c|c|cc|}
  \hline
	\multicolumn{1}{|c|}{\textbf{Name}} & \multicolumn{1}{c|}{\textbf{Scene length (s)}} & \multicolumn{2}{c|}{\textbf{Manual Rating}} \\
	 \hline 
	\multicolumn{1}{|c|}{~} & {~} & {\textbf{Valence}} & {\textbf{Arousal}}\\ \hline
	\textbf{coh}  & 127$\pm$46 & {0.08$\pm$1.18} & {1.53$\pm$0.58}\\
	\textbf{ipoh}  & 110$\pm$44 & {0.03$\pm$1.04} & {1.97$\pm$0.49}\\
	\textbf{friends}  & 119$\pm$69 & {1.08$\pm$0.37} & {2.15$\pm$0.65}\\
\hline
\end{tabular}
\vspace{-.4cm}
\label{tab:progdetails}
\end{table}

\subsection{Advertisement insertion strategy}~\label{US-AdIns}
We used three affect estimation models (audiovisual CNN, EEG CNN and manual) to provide asl, val scores for the ads and video scenes. Asl, val scores for the 24 program video scenes (8 scenes $\times$ 3 videos) were computed as mean of the ratings (in the range [-2,2] for val and [0,4] for asl) acquired from three experts, and then rescaled to [0,1] via min-max normalization. The ad affective scores were computed as follows. For the \textbf{content-centric} method, we used normalized softmax class probabilities output by the video-based CNN model~\cite{bishop:2013} for val estimation, and corresponding probabilities from the audio CNN for asl estimation. The mean score computed over all video/audio frames was used to the denote affective score of an ad in this method. Similarly, mean of the normalized softmax class probabilities over all EEG epochs for an ad was used to denote asl, val score via the user-centric \textbf{EEG} method. The average of continuous val and asl ratings in [0,1] annotated via \textit{FeelTrace}~\cite{FeelTrace} by five experts was used for \textbf{Manual} scores. 

We then adopted the CAVVA framework~\cite{cavva} to generate nine unique \textbf{video program sequences} (VPSs with average length of 19.6 minutes) with ads inserted. These VPSs represent the different combinations of the three program videos and the affect estimation approach (audiovisual/EEG/manual). Exactly five (out of seven possible) ads were inserted onto each program video. 21 of the 28 chosen ads were inserted at least once into the nine video programs, with maximum and mean insertion frequencies of 5 and 2.14 respectively. Among the 21 inserted ads, 13 had been labeled as \textit{high val} by experts, while 10 were labeled as \textit{high asl}. 

\subsection{Experiment and Questionnaire Design}~\label{US-Ques}

To evaluate the generated VPSs and thereby the efficacy of the affect estimation techniques for optimal ad insertions, we recruited 18  university undergraduates/graduates (7 female, mean age 20.1 years). Each user viewed three {VPSs} in random order such that \textit{each of the three VPSs were generated via a unique affect estimation approach}. We used a randomized 3$\times$3 Latin square design in order to cover all the nine VPSs with every three users. Thus, each VPS was seen by six of the 18 viewers, and we have a total of 54 unique user responses (18 users $\times$ three video modes per user).

We designed the user evaluation so as to reveal whether the generated VPSs (a) included seamless ad insertions, (b) facilitated user engagement towards the VPS content and (c) ensured a pleasant overall viewing experience and maximized ad memorability (both \textit{immediate} and \textit{long-term}). 

Recall evaluation is intended to verify if the inserted ads were attended to and remembered by viewers, and the immediate and day-after recall were \textit{\textbf{objective}} measures quantifying the impact of ad insertion on short-term (immediate) and long-term (day-after) memorability of the VPS-embedded ads. Specifically, we measured the proportion of (i) inserted ads that were correctly recalled (\textit{Correct} recall or \textit{hit rate}), (ii) inserted ads that were not recalled (\textit{Forgotten} or \textit{miss rate}, $ = 1-\textit{hit rate}$) and (iii) non-inserted ads incorrectly recalled as seen (\textit{Incorrect} recall or \textit{false alarm}). For those inserted ads which were correctly recalled, we also assessed whether viewers perceived them to be contextually (emotionally) relevant to the program content (\ie, whether the ad insertions were perceived to be \textit{appropriate} or \textit{good}).

Upon viewing a VPS, the viewer was provided with a representative visual frame from each of the 28 ads and a sequence-specific response sheet to test ad recall and impression concerning insertion quality. All recall and insertion quality-related responses were acquired as binary values. In addition to these objective measures, we defined a second set of \textit{\textbf{subjective}} user experience measures, and asked users to provide ratings on a 0--4 Likert scale for the questions below with 4 implying \textit{best} and 0 denoting \textit{worst}. (1) Were the ads uniformly distributed over the VPS? (2) Did the inserted ads blend well with the program flow? (3) Did the inserted ads match with the surrounding scenes in terms of \textit{content} and \textit{mood}? (4) What was the overall viewing experience while watching each VPS? Each user filled the recall and experience-related questionnaires immediately after watching each VPS. Viewers also filled in the day-after recall questionnaire, a day after completing the experiment.

\begin{figure*}[t]
\centerline{\includegraphics[width=0.33\linewidth,height=3.2cm]{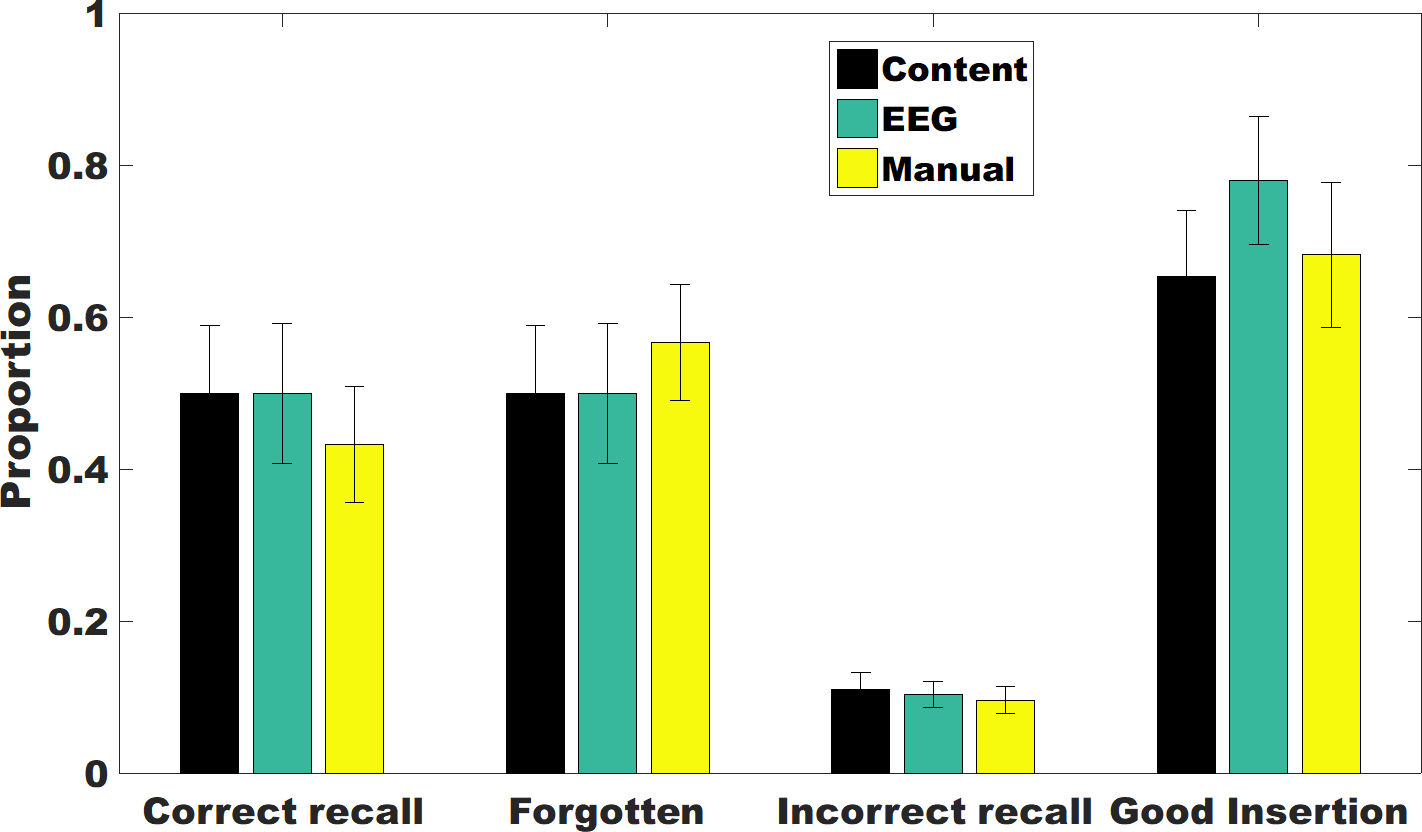}\hspace{0.05cm}
\includegraphics[width=0.33\linewidth,height=3.2cm]{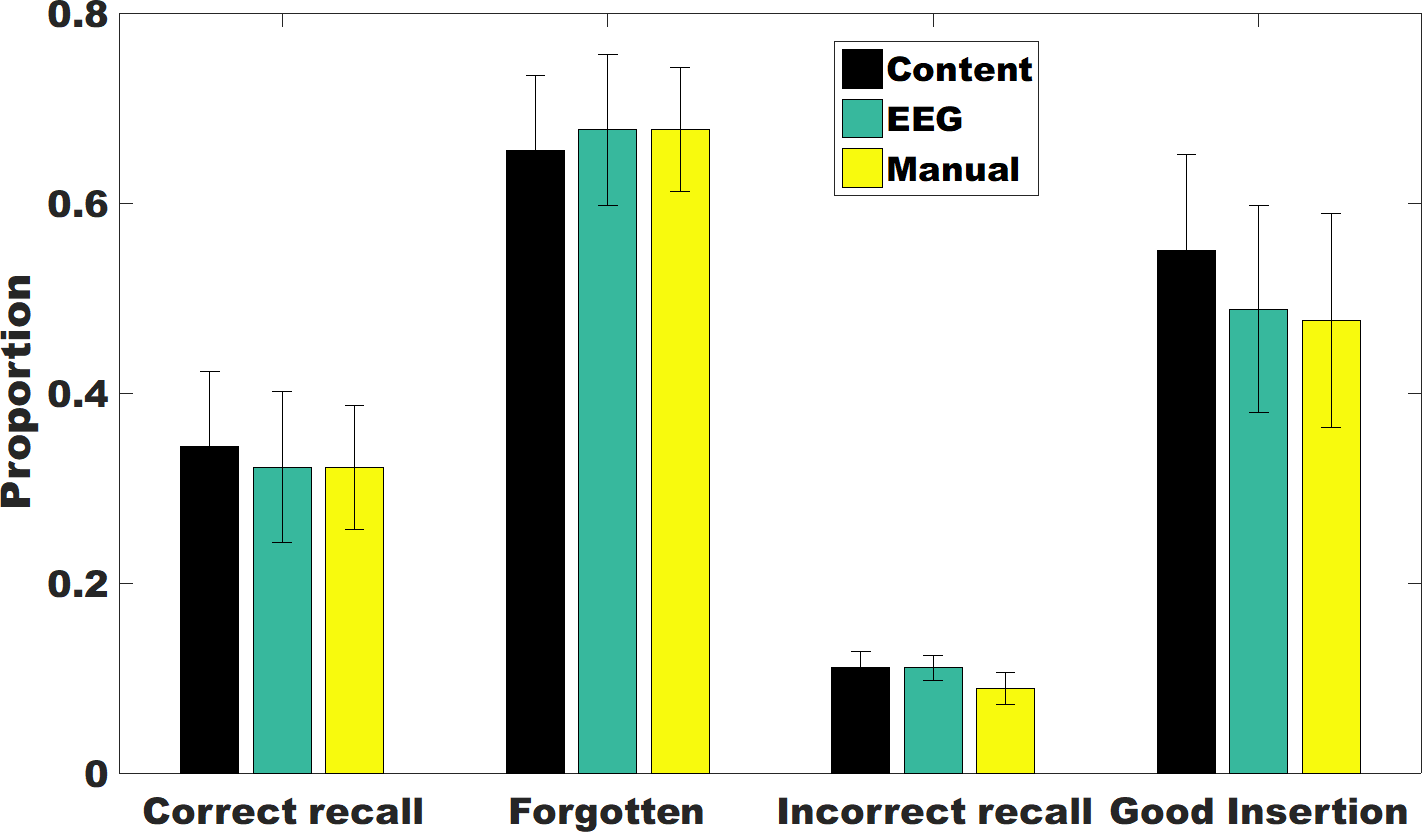}\hspace{0.05cm}
\includegraphics[width=0.33\linewidth,height=3.2cm]{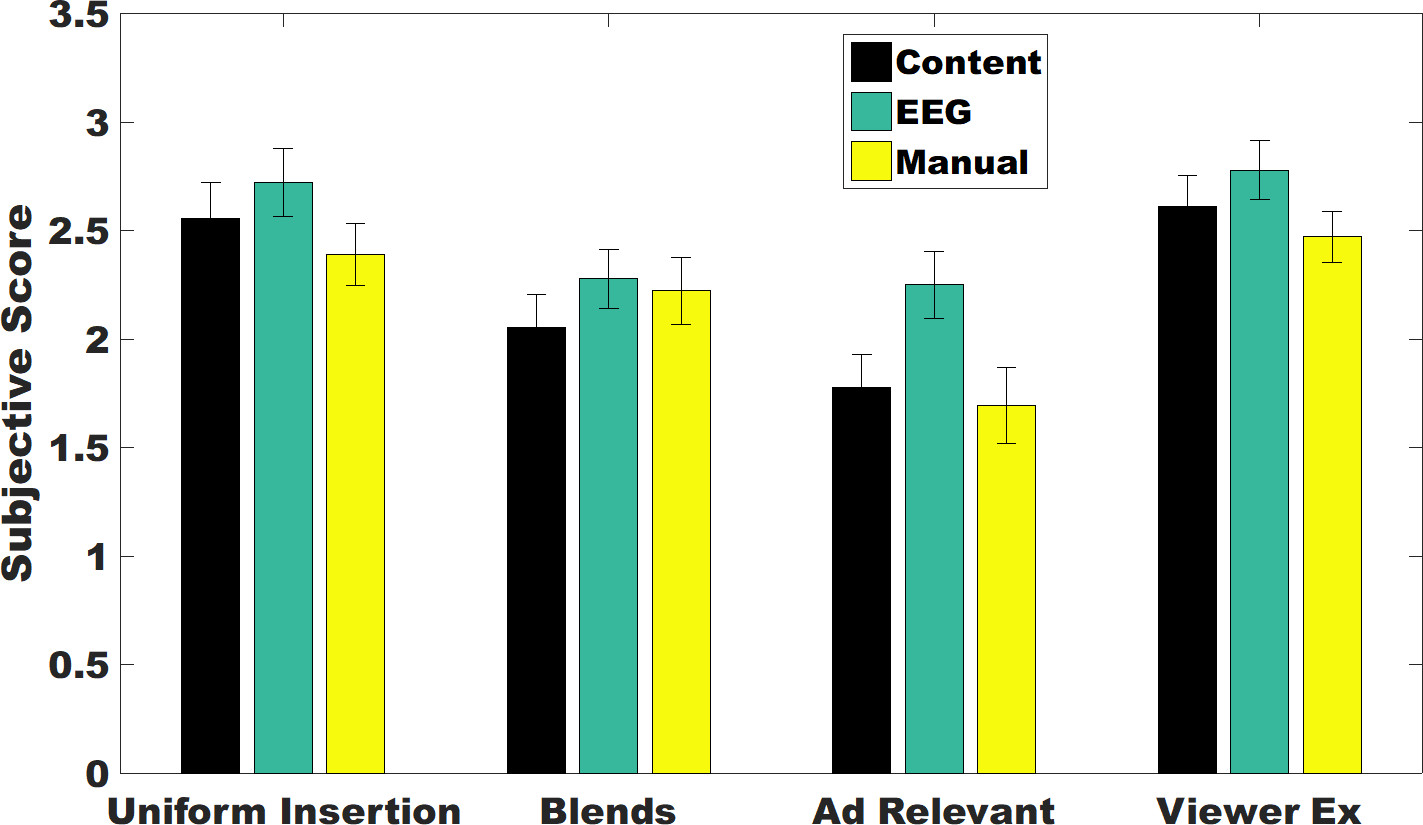}}
\centerline{\textbf{Immediate Recall}\hspace{0.2\linewidth}\textbf{Day-after recall}\hspace{0.23\linewidth}\textbf{User Experience}} \vspace{-.2cm}
\caption{\label{US_results} User study plots for recall and experience-related measures. Error bars denote standard error of mean.}
\vspace{-.5cm}
\end{figure*}

\subsection{User study results}\label{USR}
As mentioned previously, program video scenes were assigned asl, val scores manually by experts, while the content-centric CNN (denoted as `Content' hereon), EEG CNN and Manual methods were employed to estimate affective scores for ads. Overall quality of the CAVVA-generated VPS  is influenced by the quality of affective ratings assigned to both the \textit{video scenes} and \textit{ads}. In this regard, we hypothesized that better ad affect estimation would result in optimized ad insertions from the user perspective.

Firstly, we examined if there was any similarity in the ad asl and val scores estimated by the Content, EEG and Manual approaches in terms of Pearson correlations. We found that (1) there was significant and positive correlation between asl scores generated by the Manual and EEG approaches ($\rho = 0.55, p<0.005$), while asl scores computed via the Manual and Content methods ($\rho = 0.13, \text{n.s.})$ as well as via Content and EEG ($\rho = -0.22, \text{n.s.}$) were largely uncorrelated. A similar pattern was noted for val scores with a highly positive and significant correlation observed between Manual and EEG ($\rho = 0.80, p<0.000001$), while the Content--Manual ($\rho = 0.33, p=0.08$) and the Content--EEG ($\rho = 0.19, \text{n.s.}$) scores showed an insignificant positive correlation. These results are indicative of the fact that neural responses, which represent an \textit{implicit} manifestation of emotional perception/expression, best reflect \textit{explicit} affective impressions reported by humans. It is therefore unsurprising that a large number of recent affect prediction approaches~\cite{Koelstra,decaf,subramanian2016ascertain,AMIGOS} have employed neural sensing as one of the modalities incorporating emotional information.

Based on viewers' questionnaire responses, we computed the mean proportions for correct recall, ad forgottenness, incorrect recall and good insertions \textit{immediately} and a \textit{day after} the experiment. Similarly, mean subjective experience scores were computed for the three VPS generation schemes. Figure~\ref{US_results} summarizes the response results from which we make several interesting observations. 

A key measure indicative of a successful advertising strategy is \textit{high brand recall}~\cite{Holbrook1984, cavva, YadatiMMM2013}, and the immediate and day-after recall rates observed with the three ad affect estimation methods are presented in Fig.~\ref{US_results} (left),(middle). A surprising result observed from Fig.~\ref{US_results} (left) and (middle) is that ads from the content and EEG-based VPSs are better recalled (or less forgotten) than manual-based. Content-based ad insertions were best recalled both immediately and the day-after, even though recall rates for the three ad-insertion approaches were not statistically different. Given the extensive literature connecting affective attributes and memorability, we examined if any such relationships could be inferred from the user study. Overall, we found a significant and positive correlation between ad val rating and recall ($\rho = 0.44, p<0.05$) consistent with prior observations ~\cite{Subramanian2014}, in addition to the fact that about $\frac{2}{3}$rds of user-viewed ads were positive in valence.

The recall rate was much worse for the day-after condition with a high proportion of ads being forgotten. Also, the proportion of \textit{incorrectly recalled} ads was minimal in both the immediate and day-after conditions. Some discernable differences were observed in the proportion of \textit{good insertion} impressions for the three methods-- we remark here that ad recall and viewing experience are not positively correlated (some ads may be memorable because they adversely disrupted viewing experience); however, embedding ads at optimal temporal locations can enhance both ad recall and viewing experience. Post-hoc independent and right-tailed $t$-tests revealed that the proportion of immediate 'good insertion' impressions was marginally higher for EEG as compared to manual ($t_{34} = 1.337, p = 0.095$).

A number of significant differences were nevertheless observed with respect to subjective user impressions of the VPSs generated via the three methods (Fig.~\ref{US_results} (right)). The EEG-based ad insertion mechanism scored highest for all the considered criteria. Specifically, \textit{uniform insertion} scores were marginally higher for EEG with respect to manual ($t_{34} = 1.5646, p = 0.063$). A one-way balanced ANOVA on \textit{ad relevance} scores revealed the significant effect of the ad-insertion strategy ($p < 0.05$). Post-hoc $t$-tests further revealed that EEG-based ad relevance was significantly higher than manual ($t_{34} = 2.3785, p <0.05$) or content-based ($t_{34} = 2.1893, p <0.05$). EEG-based VPSs were also found to have the highest \textit{viewing experience} scores, and were significantly higher with respect to manual-based VPSs ($t_{34} = 1.7033, p <0.05$). No differences were noted with respect to user perceptions on ads \textit{blendings}.

\section{Discussion and Conclusion}~\label{CFW}

This paper discusses AR from ads, and demonstrates the utility of estimating ad asl and val (more) accurately via a computational advertising application. Firstly,  based on expert consensus we compiled a curative set of 100 semantically and emotionally diverse ads, and its ability to evoke varied-but-coherent emotions across viewers is examined by studying affective impressions of 14 raters. Suitability of the dataset for affective studies is confirmed by (1) the uniform distribution of asl and val ratings over the rating scale with minimal inter-correlation, and (2) good-to-excellent agreement between the expert and novice rater groups as measured in terms of Cohen's $\kappa$ scores.

We then evaluated the efficacy of \textit{content} and \textit{user-centric} techniques for ad AR. At the outset, we note that content and user-centered methods encode complementary emotional information. While content-centric methods examine audiovisual cues for emotion predict, they typically do not model \textit{context} which is crucial for emotion elicitation. The context may induce in the viewer an emotion very different from expectation based on the content, and therefore we hypothesized that examining user cues could be more effective as evidenced by many of the recent AR approaches.  

Our extensive content-centered AR experiments confirmed that: (1) The proposed fc7 audio and visual CNN descriptors better predicted val, and overall F1-scores revealed that video features were better at encoding emotions than spectrogram-based fc7 descriptors; (2) Multimodal methods achieved better AR than unimodal ones, and the (A+V) fc7 features produced substantially better results than audiovisual Han features for val; Probabilistic decision fusion achieved superior results with respect to feature fusion for asl, but inferior results for val.   

On the other hand, AR experiments with user-centric EEG features revealed that (1) EEG features produced superior AR performance than audiovisual descriptors; (2) The three-layer CNN classifier outperformed shallow classifiers trained on EEG data, and (3) Very comparable F1-scores were achieved with the CNN classifier with both raw (or noisy) and clean EEG data, even though shallow classifiers performed better with the cleaned features. 

The above results confirm the hypothesis that emotions are better characterized by user-centric cues, which are inherently better modulated by context~\cite{Subramanian2014} than content-centric ones. Furthermore, content-centric classification results observed over different temporal windows reveal that content features coherently reflect human impressions of val over time, but not of asl. There are two possible explanations to this end: (a) Multiple studies have found that user impressions of stimulus val are more stable and consistent as compared to asl; also the audiovisual content of ads designed to convey an element of surprise/shock is likely to exhibit significant changes over time. (b) Owing to these variations, the use of a \textit{single} affective label over the entire ad duration may be inappropriate, especially for asl, and seeking to predict time-varying affective labels could be more appropriate. Interestingly though, EEG-based AR results (Table~\ref{tab:uscap}) show only a minor deviation between the F30 and L30 conditions even for asl (lower F1-scores for the L10 condition can also be attributed to fewer training data) suggests that humans are able to grasp the general mood of advertisements fairly quickly. 

An overview of the cumulative AR results  reveals minimal model overfitting-- the variation in F1 scores across the 50 runs is fairly small in the 'All' condition for content-centric AR and over all conditions for user-centered AR. Among classifiers, RBF SVM consistently produced the best results among single-task classifiers, implying the audiovisual as well as EEG features may not be trivially linearly seperable in their respective feature spaces. However, linear multi-task learning classifier achieved close-to-ceiling performance implying that learning commonalities among similarly labeled ads facilitates better feature separability. Finally, fusing the content and user-centric results as in Table~\ref{tab:uscapfus} produced better F1-scores as compared to either modality, revealing the promise of mining both the content and user for accurate emotion prediction.

We then proceeded to check if improved emotion estimation enabled optimized ad insertion for computational advertising. Based on data compiled from 18 users, we observed that video program sequences generated via audiovisual and EEG-based affective scores were more effective in terms of ad recall and eliciting a better user experience than manually generated VPSs. Ads from content-based VPSs were recalled marginally better, both immediately and the day after. EEG-based VPSs received the highest scores for the viewing experience-related attributes. Ads in EEG-based VPSs were perceived to be (a) more uniformly distributed, and (b) more emotionally matched (or relevant) to the surrounding video scenes. Finally, EEG-based VPSs were also found to produce the best viewing experience.       

The surprising finding of audiovisual and EEG-based VPSs being superior to the manual VPS can be explained as follows. Audiovisual and EEG-based asl and val scores were estimated via CNN models, and deep CNNs have recently performed comparable to or better than humans in tasks such as object recognition~\cite{ZophVSL18} and facial expression recognition~\cite{burkert2015dexpression} due to their ability to extract fine details from data. The CAVVA optimization framework~\cite{cavva} comprises two components-- one for {selecting} \textit{ad insertion points} into the program video, and another for selecting the \textit{ads}. Asl scores only play a role in the choice of insertion points, whereas val scores influence both components. As the EEG-based framework performs best for both asl and val recognition, it also results in the most optimal ad insertions, and consequently in the best viewing experience. Finally, humans are better at rating attributes in relative than absolute terms~\cite{soleymani2008affective,AngelikiFG13}, which explains why the manually acquired ad-level asl and val scores may not be accurate (even if their general trends are consistent with the EEG scores as seen from the correlations computed in Sec.~\ref{USR}).     

The importance of context for conveying emotions via audiovisual media such as movies and ads makes context modeling critical for AR. Recurrent neural networks have shown  promise at encoding content and user-centric data for emotion~\cite{Cambria18} and mental state~\cite{BashivanRYC15} recognition. Likewise, paucity of large-scale labeled datasets in the ad AR domain motivates the use of Generative Adversarial Networks (GANs) and Variational Autoencoders (VAEs) to generate synthetic data. Future work will involve exploring these frameworks for recognizing and estimating ad emotions. 
Another line of research would be to develop algorithms that perform real-time emotional assessment of streamed video, and perform ad insertion on the fly. We will also focus on developing effective and principled methods for computational advertising, as CAVVA is modeled on \textit{ad-hoc} rules derived from consumer psychology literature. 

\section*{Acknowledgment}
This research is supported by the National Research
Foundation, Prime Ministers Office, Singapore under its
International Research Centre in Singapore Funding Initiative.

\ifCLASSOPTIONcaptionsoff
  \newpage
\fi

\bibliographystyle{IEEEtran}
\bibliography{affective_advertising_ieee}
\vspace{-.5in}


\begin{IEEEbiography}[{\includegraphics[width=1in,height=1.25in,clip,keepaspectratio]{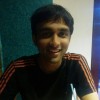}}]{Abhinav Shukla} is a Ph.D. researcher with the Imperial College, London. He was earlier a Masters student in Computer Science at the Int'l Institute of Information Technology, Hyderabad, India, from where he also received a Bachelors degree in Computer Science and Engineering. His research interests broadly lie in the fields of machine learning, computer vision and artificial intelligence.
\end{IEEEbiography}

\vspace{-.5in}

\begin{IEEEbiography}[{\includegraphics[width=1in,height=1.25in,clip,keepaspectratio]{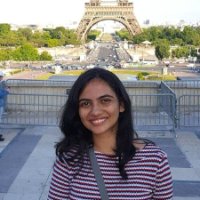}}]{Shruti Shriya Gullapuram} is a Masters student in Computer Science at the University of Massachusetts, Amherst, USA. She received her Bachelors degree in Electronics and Communication Engineering from the Int'l Institute of Information Technology, Hyderabad, India. Her research interests broadly lie in the fields of Machine Learning, Computer Vision, Human-computer Interaction and Artificial Intelligence.
\end{IEEEbiography}

\vspace{-.5in}

\begin{IEEEbiography}[{\includegraphics[width=1in,height=1.25in,clip,keepaspectratio]{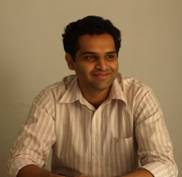}}]{Harish Katti}
received his PhD in computer science from the National University of Singapore, Masters degree in Bio-Medical Engineering from the Indian Institute of Technology, Bombay, and a B. Engg degree from Karnatak University. He worked in open standards based multimedia software development during 2000 to 2004 and was involved in the design and development of application middleware. His research interests lie broadly at the intersection of cognition and media and more specifically in experimental and computational vision research. He is currently a post-doctoral fellow at the Center for Neuroscience, Indian Institute of Science, Bangalore
\end{IEEEbiography}

\vspace{-.5in}

\begin{IEEEbiography}[{\includegraphics[width=1in,height=1.25in,clip,keepaspectratio]{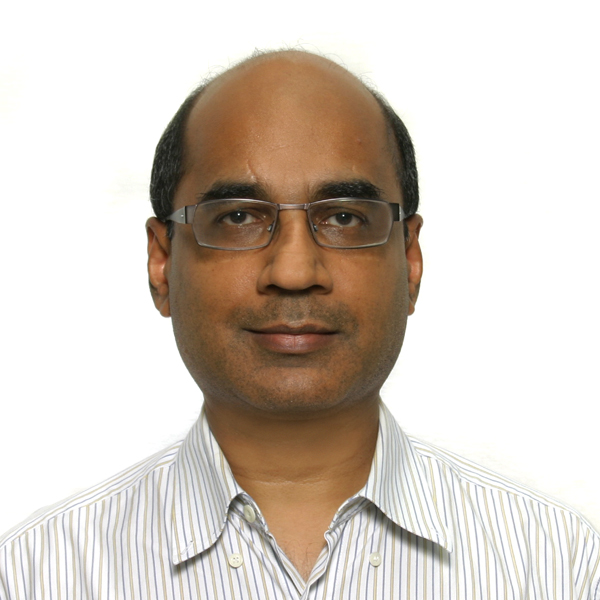}}]{Mohan Kankanhalli} is a Professor and the Dean of the School of Computing at the National University of Singapore. He earlier served as the Associate Provost for Graduate Education, the Vice-Dean for Academic Affairs and Graduate Studies and Vice-Dean for Research at the School of Computin. Mohan obtained his BTech from IIT Kharagpur and MS and PhD from the Rensselaer Polytechnic Institute. His current research interests are in Multimedia Systems (content processing, retrieval) and Multimedia Security (surveillance and privacy).  Mohan is on the editorial boards of several journals including the ACM Trans. Multimedia Computing, Communications, and Applications, Springer Multimedia Systems , Pattern Recognition and Multimedia Tools and Applications.
\end{IEEEbiography}

\vspace{-.5in}

\begin{IEEEbiography}[{\includegraphics[width=1in,height=1.25in,clip,keepaspectratio]{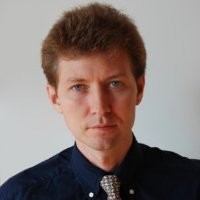}}]{Stefan Winkler} is Distinguished Scientist and Director of the Video \& Analytics Program at the University of Illinois’ Advanced Digital Sciences Center (ADSC) in Singapore. Prior to that, he co-founded a start-up,
worked for a Silicon Valley company, and held faculty positions at the National University of Singapore
and the University of Lausanne, Switzerland. He has published over 100 papers and the book “Digital Video Quality” (Wiley). He is an Associate Editor of the IEEE Transactions on Image Processing.
\end{IEEEbiography}

\vspace{-.5in}

\begin{IEEEbiography}[{\includegraphics[width=1in,height=1.25in,clip,keepaspectratio]{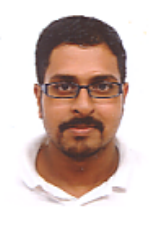}}]{Ramanathan Subramanian}
received his Ph.D. degree in Electrical and Computer engineering from  the  National University of Singapore. He is a Research Scientist at the Institute of High Performance Computing (A*STAR Singapore), and previously served as a Computer Science faculty  at the University of Glasgow (Singapore) and the Int'l  Institute  of  Information  Technology, Hyderabad (India). His research focuses on Human-centered and Human-assisted computing, and specifically on applications which utilize non-verbal human behavioral cues for media and user analytics. He is a Senior Member of IEEE and a member of the ACM and AAAC.
\end{IEEEbiography}

\end{document}